\documentclass[journal]{IEEEtran}
\usepackage[tbtags]{amsmath}
\usepackage{amsfonts}
\usepackage{amssymb}
\usepackage{subfigure}
\usepackage{cite}
\usepackage{calc}
\usepackage{color}
\usepackage{epsfig}
\usepackage{setspace}

\newtheorem{defn}{Definition}
\newtheorem{thm}{Theorem}[section]
\newtheorem{cor}[thm]{Corollary}
\newtheorem{prop}{Proposition}
\newtheorem{lem}[thm]{Lemma}
\newtheorem{conj}[thm]{Conjecture}
\newtheorem{constr}[thm]{Construction}
\newtheorem{note}{Remark}
\newcommand{\bit}{\begin{itemize}}
\newcommand{\eit}{\end{itemize}}
\newcommand{\bcor}{\begin{cor}}
\newcommand{\ecor}{\end{cor}}
\newcommand{\beq}{\begin{equation}}
\newcommand{\eeq}{\end{equation}}
\newcommand{\beqn}{\begin{equation*}}
\newcommand{\eeqn}{\end{equation*}}
\newcommand{\beqa}{\begin{eqnarray}}
\newcommand{\eeqa}{\end{eqnarray}}
\newcommand{\beqan}{\begin{eqnarray*}}
\newcommand{\eeqan}{\end{eqnarray*}}
\newcommand{\ben}{\begin{enumerate}}
\newcommand{\een}{\end{enumerate}}
\newcommand{\bdefn}{\begin{defn}}
\newcommand{\edefn}{\end{defn}}
\newcommand{\bnote}{\begin{note}}
\newcommand{\enote}{\end{note}}
\newcommand{\bprop}{\begin{prop}}
\newcommand{\eprop}{\end{prop}}
\newcommand{\blem}{\begin{lem}}
\newcommand{\elem}{\end{lem}}
\newcommand{\bthm}{\begin{thm}}
\newcommand{\ethm}{\end{thm}}
\newcommand{\bconj}{\begin{conj}}
\newcommand{\econj}{\end{conj}}
\newcommand{\bconstr}{\begin{constr}}
\newcommand{\econstr}{\end{constr}}
\newcommand{\bpf}{\begin{proof}}
\newcommand{\epf}{\end{proof}}

\begin{document}

\title{DMT of Multi-hop Cooperative Networks - Part I: Basic Results}

\author{\authorblockN{K. Sreeram, S. Birenjith, and P. Vijay Kumar} \\
\thanks{The authors are with the Department of Electrical Communication Engineering, Indian Institute of Science, Bangalore, India
(Email: sreeramkannan@ece.iisc.ernet.in, biren@ece.iisc.ernet.in,
vijayk@usc.edu). P.~Vijay Kumar is on leave of absence from the
University of Southern California, Los Angeles, USA. }
\thanks{This work was supported in part by NSF-ITR Grant CCR-0326628, in part by the DRDO-IISc Program on Advanced
Mathematical Engineering and in part by Motorola's University
Research Partnership Program with IISc.}
\thanks{The material in this paper was presented in part at the 10th International
Symposium on Wireless Personal Multimedia Communications, Jaipur,
Dec. 2007, the Information Theory and Applications Workshop, San
Diego, Jan. 2008, and at the IEEE International Symposium on
Information Theory, Toronto, July 2008.} }

\date{}
\maketitle

\begin{abstract}

In this two-part paper, the DMT of cooperative multi-hop networks
is examined. The focus is on single-source single-sink (ss-ss)
multi-hop relay networks having slow-fading links and relays that
potentially possess multiple antennas. In this first part, some
basic results that help in determining the DMT of cooperative
networks as well as in characterizing the two end-points of the
DMT for arbitrary full-duplex networks is established. In the
companion paper, two families of half-duplex networks are studied.

The present paper examines the two end-points of the DMT of ss-ss
networks. In particular, the maximum achievable diversity of
arbitrary multi-terminal wireless networks is shown to be equal to
the min-cut between the corresponding source and the sink. The
maximum multiplexing gain (MMG) of arbitrary full-duplex ss-ss
networks is shown to be equal to the min-cut rank, using a new
connection to a deterministic network for which the capacity was
recently found. This connection is operational in the sense that a
capacity-achieving scheme for the deterministic network can be
converted into a MMG-achieving scheme for the original network.

We also prove some basic results including a proof that the
colored noise encountered in AF protocols for cooperative networks
can be treated as white noise for DMT computations. We derive
lower bounds for the DMT of triangular channel matrices, which are
subsequently utilized to derive alternative, and often simpler
proofs of several existing results. The DMT of a parallel channel
with independent MIMO links is also computed here. As an
application of these basic results, we prove that a linear
tradeoff between maximum diversity and maximum multiplexing gain
is achievable for arbitrary, ss-ss single-antenna,
directed-acyclic networks equipped with full-duplex relays.

All protocols in this paper are explicit and rely only upon
amplify-and-forward (AF) relaying. Explicit codes for all
protocols introduced here are included in the companion paper.

\end{abstract}

\section{Introduction\label{sec:introduction}}

In fading relay networks, cooperative diversity provides a means
of operating the network efficiently.  While much of the work in the
literature on cooperative diversity is based on two-hop networks,
the attention here is on multi-hop networks.

\subsection{Prior Work\label{sec:prior_work}}

The concept of user cooperative diversity was introduced in
\cite{SenErkAaz1}. Cooperative diversity protocols were first
discussed in \cite{LanWor} for the two-hop, single-relay network
(Fig.\ref{fig:classical_relay}). Zheng and Tse \cite{ZheTse}
proposed the Diversity-Multiplexing gain Tradeoff (DMT) as a means
of evaluating point-to-point, multiple-antenna schemes in the
context of slow-fading channels.

\subsubsection{Two-hop Networks}
\begin{figure}[h]
  \centering
  \subfigure[General two-hop relay network]{\label{fig:classical_relay}\includegraphics[height=30mm]{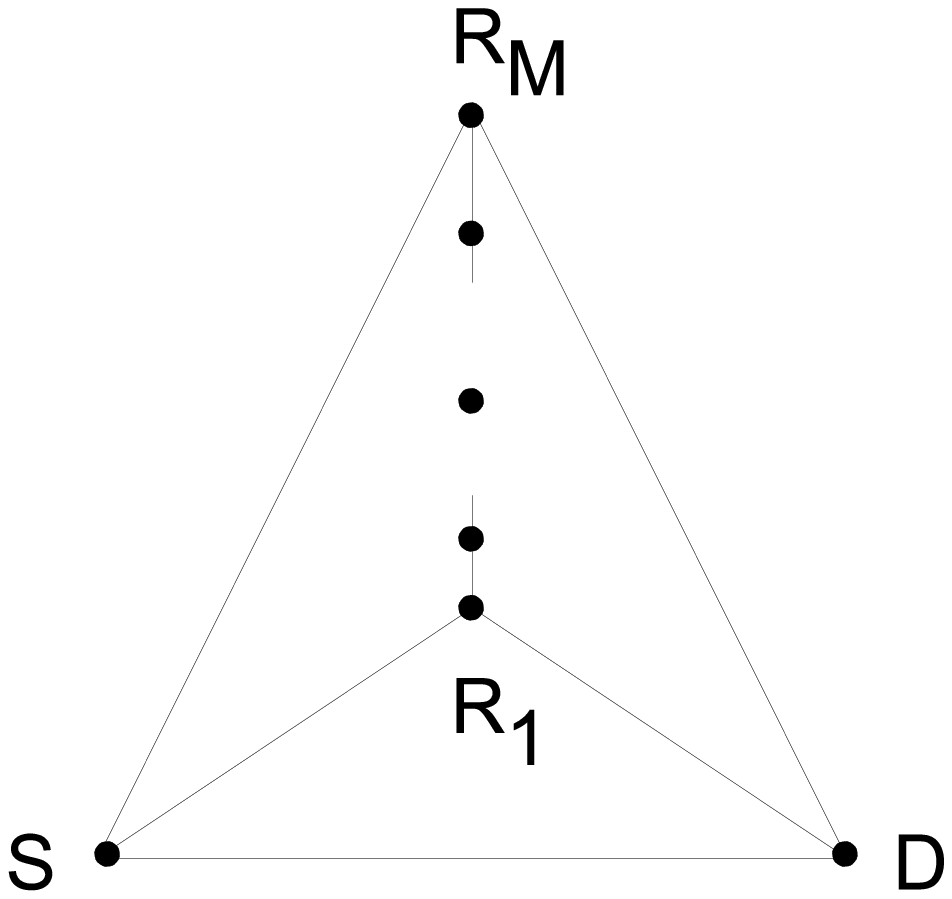}}
  \subfigure[Single relay network]{\label{fig:one_relay}\includegraphics[width=40mm]{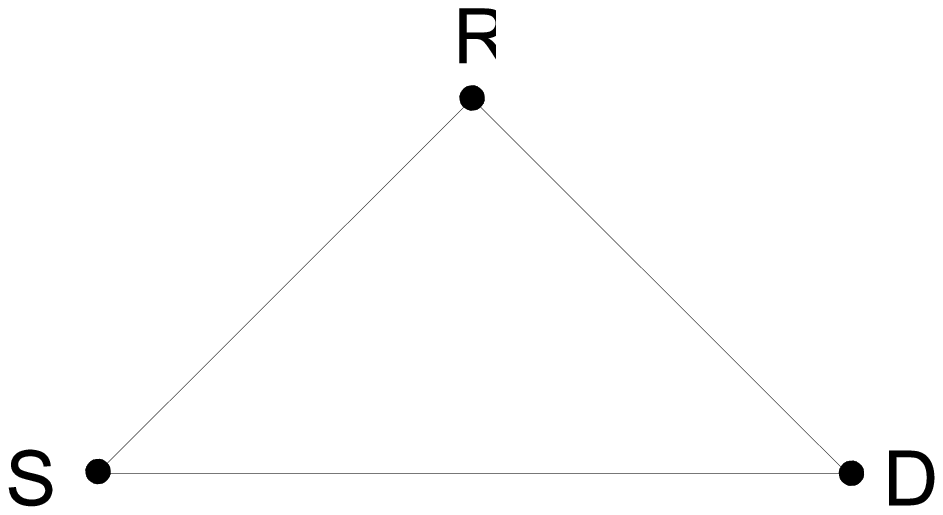}}
  \caption{Two-hop cooperative relay networks}
  \label{fig:two_hop_relay}
\end{figure}

The DMT was also used as a tool to compare various protocols for
half-duplex two-hop cooperative networks in
\cite{LanTseWor,AzaGamSch}. As noted in \cite{YukErk}, the DMT is
simple enough to be analytically tractable and powerful enough to
compare different cooperative-relay-network protocols. For any
network, an upper bound on the achievable DMT is given by the
cut-set bound \cite{CovTho,YukErk}. A fundamental question in this
area is whether the cut-set bound on DMT can be achieved. While
this question has been studied extensively for the two-hop
cooperative wireless system in Fig.\ref{fig:classical_relay}, the
question still remains open even for this class of network (see
\cite{EliVinAnaKum}, \cite{PraVar} for a detailed comparison of
existing achievable regions).

In \cite{LanTseWor}, the selection-decode-and-forward protocol is
analyzed for an arbitrary number of relays, where the authors give
upper and lower bounds on the DMT of the protocol. In these
protocols, the relays and the source node participate for equal
time instants and the maximum multiplexing gain $r$ achieved is
equal to $0.5$.

In \cite{AzaGamSch}, Azarian {\it et al.} analyze the class of Non
Orthogonal, Amplify-and-Forward (NAF) protocols, introduced earlier
by Nabar {\it et al.} in \cite{NabBolKne} and establish the improved DMT of the NAF protocol in
comparison to the class of Orthogonal-Amplify-and-Forward (OAF)
protocols considered in \cite{LanTseWor}. It has been shown in
\cite{EliVinAnaKum}, that the DMT of the NAF protocol can be
obtained via the OAF protocol as well using appropriate unequal
slot lengths for source and relay transmissions.

The authors of \cite{AzaGamSch} also introduce the Dynamic
Decode-and-Forward (DDF) protocol wherein the time duration for
which the relays listen to the source depends on the source-relay
channel gain. They show that for the single-relay case, the DMT of
the DDF protocol achieves the cut-set bound (also known as the
transmit-diversity bound) for $r \leq 0.5$, beyond which the DMT
falls below the bound. An enhanced DDF protocol is proposed in
\cite{PraVar} that improves upon DDF. However the DMT of this
protocol also falls short of the transmit bound for $r \geq 0.5$.

Yang and Belfiore consider a class of protocols called
Slotted-Amplify-And-Forward (SAF) protocols in \cite{YanBelSaf}
for the two-hop network with direct link, and show that these
improve upon the performance of the NAF protocol \cite{AzaGamSch}
for the case of two relays. Under the assumption of relay
isolation, the naive SAF scheme proposed in \cite{YanBelSaf} is
shown to achieve the cut-set bound. It is also conjectured in
\cite{YanBelSaf} that SAF protocol is optimal even when the relays
are not isolated.

Yuksel and Erkip in \cite{YukErk} have considered the DMT of the
DF and compress-and-forward (CF) protocols. They show that the CF
protocol achieves the transmit-diversity bound for the case of a
single relay. We note, however, that in the CF protocol, the
relays are assumed to know all the fading coefficients in the
system. The authors also translate cut-set upper bounds in
\cite{CovTho} for mutual information into the DMT framework for a
general multi-terminal network.

Jing and Hassibi \cite{JinHas} consider cooperative communication
protocols for the two-hop network without a direct link between
source and destination. They study protocols where the relay nodes
apply a linear transformation to the received signal and analyze
their BER performance. The authors consider the case when the
source and the relays transmit for an equal number of channel uses
and the relays perform a unitary transformation on the input
symbols before transmitting it. Rao and Hassibi \cite{RaoHas}
consider two-hop half-duplex multi-antenna cooperative networks
without direct link and provide an AF scheme and compute the DMT
achieved by the scheme. Their scheme incurs a rate loss of a
factor of two compared to the cut-set bound. In a parallel work
\cite{GhaBayKha}, the DMT of the two-hop network without direct
link is proved to be equal to the cut-set bound.

\subsubsection{Multi-hop Networks} Yang and Belfiore in
\cite{YanBelNew} consider AF protocols for a family of MIMO
multi-hop networks (which are termed as multi-antenna layered
networks in the current paper). They derive the optimal DMT for
the Rayleigh-product channel which they prove is equal to the DMT
of the AF protocol applied to this channel. They also propose AF
protocols to achieve the optimal diversity of these multi-antenna
layered networks.

Oggier and Hassibi \cite{OggHas} have proposed distributed space
time codes for multi-antenna layered networks that achieve
diversity gain equal to the minimum number of relay nodes among
the hops. Recently, Vaze and Heath \cite{VazHea} have constructed
distributed space time codes based on orthogonal designs that
achieve the optimal diversity of the multi-antenna layered network
with low decoding complexity. In \cite{VazHea2}, the same authors
study the circumstances under which full diversity can be achieved
without coding in a layered network in the presence of partial
CSIT.

Borade, Zheng and Gallager in \cite{BorZheGal} consider AF schemes
on a class of multi-hop layered networks where each layer has the
same number of relays (termed as regular networks in the current
paper). They show that AF strategies are optimal in terms of
multiplexing gain. They also compute lower bounds on the DMT of
the product Rayleigh channel.

\subsubsection{Capacity} There has been a recent interest in determining
approximations to the capacity of wireless networks. The pre-log
coefficient of the capacity, termed as the degrees-of-freedom
(DOF) of wireless multi-antenna networks is studied in
\cite{GodHero} \cite{HostNosr}. \footnote{The degrees-of-freedom
is alternately referred to as maximum multiplexing gain in the
literature, although the former is typically used for ergodic
capacity characterizations and the latter is typically used in the
context of outage characterization. This paper deals with the DMT,
which is a outage characterization and for this reason, we use the
term multiplexing gain.} The DOF for the $N$ user interference
channel was derived in \cite{CadJaf}, for the MIMO X networks in
\cite{JafSha,MadMotKha}  and the DOF of single-source single-sink
(ss-ss) layered networks was obtained in \cite{BorZheGal}.

In a different direction, the capacity of ss-ss and multi-cast
deterministic wireless networks has been characterized in
\cite{AveDigTse1}.

Intuition drawn from the deterministic wireless networks was used
to identify capacity to within a constant for some example
networks in \cite{AveDigTse2}. Very recently, the capacity of
single-antenna gaussian relay networks has been characterized to
within a constant number of bits in \cite{AveDigTse3}. This result
also easily extends to give the approximate compound-channel
capacity for full-duplex single-antenna networks. The results in
\cite{AveDigTse3} can also be used to show that for half duplex
networks, under any fixed schedule of operation, the best possible
rate can be achieved (to within a constant number of bits).
However the determination of optimal schedules that achieve the
maximum possible DMT remains open, which we solve for certain
classes of networks in the companion paper.

In \cite{NazGas}, given a wireline network code, a scheme for
wireless gaussian relay channel is obtained where each relay
computes linear transformations of its input signals and the
achievable rate region for the scheme is characterized.

\subsubsection{Codes}  Cyclic Division Algebras (CDA) were first
used to construct space-time codes in \cite{SetRajSas}.  The
notion of space-time codes having a non-vanishing determinant
(NVD) was introduced in \cite{BelRekVit}. Subsequently, it was
shown in \cite{EliRajPawKumLu} that CDA-based ST codes with NVD
achieve the DMT of the Rayleigh-fading channel and minimal-delay
codes with NVD were constructed for all $n_t$. From the results in
\cite{TavVis}, these codes are moreover, approximately universal,
i.e., DMT optimal for every statistical characterization of the
fading channel.

These codes were tailored to suit the structure of various static
protocols for two-hop cooperation and proved to be DMT optimal for
certain protocols in \cite{EliVinAnaKum}. For the DDF protocol,
DMT optimal codes were constructed for arbitrary number of relays
with multiple antennas in \cite{EliKum}. For the specific case of
single-relay single-antenna DDF channel, codes were constructed
recently in \cite{RajCaire}, which are not only DMT optimal, but
also have probability of error close to the outage probability.
Codes for the multi-antenna two-hop network under the NAF protocol
were presented in \cite{YanBelMimoAf}. CDA-based ST codes
construction for the rayleigh parallel channel were provided in
\cite{Lu,YanBelRek}.  This construction was shown to be
approximately universal for the class of MIMO parallel channels in
\cite{EliKum}. In this paper, we present a DMT optimal code design
for all proposed protocols based on the approximately universal
codes in \cite{EliRajPawKumLu} and \cite{EliKum}.

\subsubsection{Other Work} Cooperative networks with asynchronous transmissions
have also been studied in the literature
\cite{Wei},\cite{LiXia},\cite{RajRajDist}. However, we consider
networks in which relays are synchronized. Codes for two-hop
cooperative networks having low decoding complexity and full
diversity are studied in \cite{JinJaf}, \cite{RajRajDist} and
\cite{YiKim}. While decoding complexity is not the primary focus
of the present paper, we do provide a
successive-interference-cancellation technique to reduce the code
length and thereby, the complexity.

\subsection{Setting and Channel Model \label{sec:channel_model}}

\subsubsection{Network Representation by a Graph}
Unless otherwise stated, all networks considered possess a single
source and a single sink and we will apply the abbreviation ss-ss
to denote these networks. Any wireless network can be associated
with a directed graph, with vertices representing nodes in the
network and edges representing connectivity between nodes. If an
edge is bidirectional, we will represent it by two edges, one
pointing in either direction. An edge in a directed graph is said
to be \emph{live} at a particular time instant if the node at the
head of the edge is transmitting at that instant. An edge in a
directed graph is said to be \emph{active} at a particular time
instant if the node at the head of the edge is transmitting and
the tail of the edge is receiving at that instant.

A wireless network is characterized by broadcast and interference
constraints. Under the \emph{broadcast} constraint, all edges
connected to a transmitting node are simultaneously live and
transmit the same information. Under the \emph{interference}
constraint, the symbol received by a receiving end is equal to the
sum of the symbols transmitted on all incoming live edges. We say
that a protocol avoids interference if only one incoming edge is
live for all receiving nodes.

In wireless networks, the relay nodes operate in either half or
full-duplex mode. In case of half-duplex operation, a node cannot
simultaneously listen and transmit, i.e., an incoming edge and an
outgoing edge of a node cannot simultaneously be active.

In this paper, we use uppercase letters to denote matrices and
lowercase letters to denote vectors/scalars. Vectors and scalars
are differentiated only through the context. Irrespective of
whether a particular random entity is a scalar, vector or a
matrix, the entity will be represented using boldface letters.

Between any two adjacent nodes $v_x$, $v_y$ of a wireless network, we assume the following channel model.

\beq
    \bold{y} \ = \bold{ H {x} + {w}} \ ,
    \label{eq:channel_model}
\eeq where $\bold{y}$ corresponds to the received signal at node $v_y$, $\bold{w}$ is the noise vector,
$\bold{H}$ is a matrix and $\bold{x}$ is the vector transmitted by the node $v_x$.

\subsubsection{Assumptions} We follow the literature in making the assumptions
listed below. Our description is in terms of the equivalent complex-baseband, discrete-time channel.

\ben \item All channels are assumed to be quasi-static and to
experience Rayleigh fading and hence all fade coefficients are
i.i.d., circularly-symmetric complex gaussian $\mathbb{C}\mathcal
{N} (0,1)$ random variables. \item The additive noise at each
receiver is also modelled as possessing an i.i.d.,
circularly-symmetric complex gaussian $\mathbb{C}\mathcal {N}
(0,1)$ distribution. \item Each receiver (but none of the
transmitters) is assumed to have perfect channel state information
of all the upstream channels in the network. \footnote{However,
for the protocols proposed in this paper, the CSIR is utilized
only at the sink, since all the relay nodes are required to simply
amplify and forward the received signal.} \een

\subsection{Results \label{sec:results}}

In this paper, we characterize maximum diversity, maximum
multiplexing gain and achievable DMT for arbitrary cooperative
networks. Some of these results were presented in conference
versions of this paper \cite{WPMC,ITA,ISIT1,ISIT3} (see also
\cite{Arxiv,TechReport}). Special classes of networks are
considered in the second part of this two-part paper,
\cite{Part2}. Optimal code design for all proposed protocols in
both parts of the paper can also be found there.

The principal results established in this paper are the following
(see Table~\ref{tab:summary} for a tabular of results). \ben \item
The maximum diversity of a multi-antenna multi-terminal network is
equal to the value of the min-cut between the source and the
destination. \item The maximum multiplexing gain for a ss-ss
full-duplex multi-antenna network is equal to the minimum rank of
any cut between the source and the destination. \item A DMT which
is linear between the maximum diversity and maximum multiplexing
gain is achievable for full-duplex single-antenna relay networks.
\een

We also prove the following general results, that are useful in
computing the DMT of cooperative networks \ben
\setcounter{enumi}{3} \item The colored noise encountered in
cooperative networks can be treated as white for DMT computations.
\item We provide a lower bound on the DMT of triangular matrices.
\item We compute the DMT of a parallel MIMO channel in terms of
the DMT of the component MIMO links. \een

\begin{table*}
\label{tab:summary} \caption{Principal Results Summary}
\begin{center}
\begin{tabular}{||c|c|c|c|c|c|c|c|c||}
\hline \hline &&&&&&&&\\
Network  & No of    & No of     & FD/ & Direct & Upper bound on  & Achievable & Is upper bound& Reference  \\
         & sources/ & antennas  & HD  &  Link  & Diversity/DMT   & Diversity/DMT& achieved? &\\
         &  sinks   & in nodes  &        &        & $d_\text{bound}(r)$  & $d_\text{achieved}(r)$& &\\
\hline \hline &&&&&&&&\\
Arbitrary      & Multiple     & Multiple     & FD/HD & $\checkmark$  & $d(0)=$ Min-cut & $d(0)=$  Min-cut & $\checkmark$&Theorem~\ref{thm:mincut}\\
&&&&&&&($d_{\max}$ achieved)&\\
&&&&&&&&\\
\hline
&&&&&&&&\\
Arbitrary     & Multiple & Multiple   & FD/HD & $\times$ & $d(0)=$ Min-cut & $d(0)=$ Min-cut & $\checkmark$&Theorem~\ref{thm:mincut}\\
&&&&&&&($d_{\max}$ achieved)&\\
&&&&&&&&\\
\hline
&&&&&&&&\\
Arbitrary & Single  & Multiple    & FD & $\checkmark$ & $r_{\max}$ = Rank of & $r_{{\max}}$ = Rank of &$\checkmark$ &Theorem~\ref{thm:DOF_ss}\\
&&&&& Min-cut &Min-cut&($r_{max}$ achieved)&\\
&&&&&&&&\\
\hline
&&&&&&&&\\
Arbitrary & Single  & Single    & FD & $\checkmark$ & Concave & $d_{\max}(1-r)^+$ & A linear DMT&Theorem~\ref{thm:FD_No_Direct_Path}\\
Directed &&&&&in general&& between $d_{\max}$ and&\\
Acyclic&&&&&&& $r_{\max}$ is achieved&\\
Networks&&&&&&&&\\
&&&&&&&&\\
\hline \hline
\end{tabular}
\end{center}
\end{table*}

\subsection{Relation to Existing Literature}

\ben \item \emph{Proof of a Conjecture by Rao and
Hassibi:}

The results in \emph{Example 5} in
Section~\ref{sec:examples_main_thm} prove Conjecture 1 of
\cite{RaoHas} and \cite{Rao}.

\item \emph{Lower bound on the DMT of various AF Protocols:}
Certain results in this paper can be used to recover existing
results on the DMT of AF protocols in a simpler, concise and more
intuitive manner.

\emph{NAF Protocol:} We compute a lower bound on the DMT of the
NAF protocol, which turns out to be tight, as proved in
\cite{AzaGamSch}.

\emph{SAF Protocol:} We compute a lower bound on the DMT of the
Slotted Amplify-and-Forward (SAF) protocol under the
relay-isolation assumption \cite{YanBelSaf} in \emph{Example 2} of
Section~\ref{sec:examples_main_thm}.  From the results in
\cite{YanBelSaf}, this lower bound is in fact tight.

\emph{N-Relay MIMO NAF Channel Appearing in \cite{YanBelMimoAf}:}

In \emph{Example 5} of Section~\ref{sec:examples_main_thm}, we prove an improved lower bound on the DMT for the
MIMO NAF protocol for a two-hop multi-antenna network with a direct link compared to the bound in
\cite{YanBelMimoAf}.

\item \emph{The diversity of arbitrary cooperative networks.}

As noted earlier, we characterize completely the maximum diversity
order attainable for arbitrary cooperative networks and it is
shown that an amplify-and-forward scheme is sufficient to achieve
this. Special cases of these were derived for the MIMO two-hop
relay channel in \cite{YanBelMimoAf}, under a certain condition on
the number of antennas (See \ \emph{Corollary 1} in that paper).
Also, the diversity order of layered networks using
amplify-and-forward networks is characterized in \cite{YanBelNew}.
The same result is obtained using lower-complexity codes in
\cite{VazHea} and \cite{VazHea2}. For arbitrary ss-ss networks,
upper bounds on the diversity order of ss-ss networks are derived
in \cite{BoyFalYan}, however, no achievability results are given
there. Very recently, \cite{GhaBayKha} have characterized the
diversity of general ss-ss networks. It must be noted that this
result can be obtained as a special case of our result for
multi-terminal networks, which appeared in \cite{ITA}, although
the achievability strategy is different in \cite{GhaBayKha}.

\item \emph{DMT of single-antenna full-duplex networks} As a
consequence of the compound channel results in \cite{AveDigTse3},
the optimal DMT of full-duplex single-antenna networks can be
proved to be equal to the cut-set bound. While most of the results
in the current paper focus on either multi-antenna or half-duplex
networks, it must be noted that the schemes presented in
\cite{AveDigTse3} involve long random coding arguments in contrast
to the short block-length, explicit schemes presented in the
present paper.

\item \emph{Maximum multiplexing gain of cooperative relay
networks} The maximum multiplexing gain for single-antenna
full-duplex relay networks can be readily obtained from the
results in \cite{AveDigTse3} and it is potentially possible to
extend these results to the multiple antenna case.  We adopt
however, a different approach here, and utilize a conversion from
the deterministic wireless network to the fading network in order
to determine the MMG. The conversion is operational in the sense
that a capacity achieving strategy on deterministic network can be
converted into a MMG-achieving strategy for the fading network.

\item  The DMT of the parallel channel in closed form is obtained
in Lemma~\ref{lem:parallel_channel}. A special case of this result is derived in \cite{YanBelNew} where the
authors characterize the parallel channel DMT for the case when all the individual channels have the same DMT.
\een

\subsection{Outline \label{sec:outline}}

In Section \ref{sec:general_theory}, we present basic results and
techniques which will be of use in studying the DMT of multi-hop
networks. In this section, we introduce the information-flow
diagram (i-f diagram), and prove a lower bound on the DMT of lower
triangular matrices. In Section~\ref{sec:extreme_points}, we
characterize the extreme points of the optimal DMT of arbitrary
ss-ss networks. We provide a lower bound to the DMT of arbitrary
ss-ss networks with single-antenna, full-duplex relays in
Section~\ref{sec:full_duplex}.

In the sequel to the present paper, we will make use of the basic
results and techniques introduced here, to characterize the
optimal DMT of certain classes of networks. The second part will
also provide code designs for all the protocols proposed in both
parts of the paper.

\section{Basic Results for Cooperative Networks \label{sec:general_theory}}

We begin by reviewing the notion of DMT in point-to-point channels
and then go on to explain how the DMT becomes a meaningful tool in
the study of cooperative wireless networks. Later in this section,
we develop general techniques, which will prove useful in deriving
results on the optimal DMT of ss-ss networks.

\subsection{Background}

\subsubsection{Diversity-Multiplexing Gain Tradeoff\label{sec:dmt}}

Let $R$ denote the rate of communication across the network in
bits per network use. Let $\wp$ denote the protocol used across
the network, not necessarily an AF protocol.  Let $r$ denote the
multiplexing gain associated to rate $R$ defined by \beqan R & = &
r \log (\rho) .\eeqan  The probability of outage for the network
operating under protocol $\wp$, i.e., the probability of outage of
the induced channel in \eqref{eq:channel_model} is then given by
\beqan
    \lefteqn{P_{\text{out}}(\wp,R) = } \\
     & & \inf_{\Sigma_x \ \geq \ 0, \
    \text{Tr} \ (\Sigma_x) \ \leq \ n \rho }
    \Pr( I(\bold{x};\bold{y}) \ < \ n R | \bold{H(\wp)} = H(\wp) ),
\eeqan where $\bold{H(\wp)}$ denotes the collection of all random
variables associated with the induced channel of the protocol
$\wp$. Let the outage exponent $d_{\text{out}}(\wp,r)$ be defined
by \beqan d_{\text{out}}(\wp,r) & = & - \lim_{\rho \rightarrow
\infty} \frac{ P_{\text{out}}(\wp,R)}{\log( \rho) }, \eeqan and we
will indicate this by writing \beqan \rho^{-d_{\text{out}}(\wp,r)}
& \doteq & P_{\text{out}}(\wp,R). \eeqan  The symbols $\dot \geq$,
$\dot \leq$ are similarly defined.

The outage exponent $d_{\text{out}}(r)$ of the network associated
to multiplexing gain $r$ is then defined as the supremum of the
outages taken over all possible protocols, i.e., \beqan
d_{\text{out}}(r) & = & \sup_{\wp} d_{\text{out}}(\wp,r). \eeqan

A distributed space-time code (more simply, a code) operating
under a protocol $\wp$ is said to achieve a diversity gain
$d(\wp,r)$ if \beqn P_{e}(\wp,\rho) \doteq \rho^{-d(\wp,r)} \ ,
\eeqn where $P_e(\rho)$ is the average error probability of the
code $C(\rho)$ under maximum likelihood decoding. Using Fano's
inequality, it can be shown (see \cite{ZheTse}) that for a given
protocol, \beqan
    d(\wp,r) & \leq & d_{\text{out}}(\wp,r).
\eeqan

The DMT $d(r)$ of the network associated to a multiplexing gain
$r$ is then defined as the supremum of all achievable diversity
gains across all possible protocols and codes.

We will refer to the outage exponent $d_{\text{out}}(r)$ of a
protocol in this paper as the DMT $d(r)$ of the protocol, since
for every protocol discussed in this paper, we shall identify a
corresponding coding strategy that achieves $d(\wp,r)$ in the
sequel \cite{Part2} to the present paper.

\bdefn \label{defn:DMT_Matrix} Given a random matrix $\bold{H}$ of
size $m \times n$, we define the \emph{DMT of the matrix}
$\bold{H}$ as the DMT of the associated channel $\bold{y = Hx +
w}$ where $\bold{x}$ and $\bold{y}$ are column vectors of size $(n
\times 1)$ and $(m \times 1)$ respectively, and where $\bold{w}$
is a $\mathcal{CN}(0,I)$ column vector. We denote the DMT of the
matrix $\bold{H}$ by $d_H(.)$ \edefn

\subsubsection{Cut-Set bound on DMT} On any network,
the cut-set upper-bound on mutual information of a general
multi-terminal network~\cite{CovTho} translates into an upper
bound on the DMT. This was formalized in \cite{YukErk} as follows:

\blem \label{lem:CutsetUpperBound} Let $r \log(\rho)$ be the rate
of communication between the source and the sink. Given a cut
$\omega$ between source and destination, let $\bold{H}_{\omega}$
denote the transfer matrix between nodes on the source-side of the
cut and those on the sink-side, and let $d_{\omega}(r)$ be the DMT
of $\bold{H}_{\omega}$. Then the DMT of communication between
source and destination is upper bounded by \beqn d({r}) \leq
\min_{\omega \in \Lambda} \{d_{\omega}(r)\}, \eeqn where $\Lambda$
is the set of all cuts between the source and the destination.
\elem

\begin{figure}[h!]
\centering
\includegraphics[width=70mm]{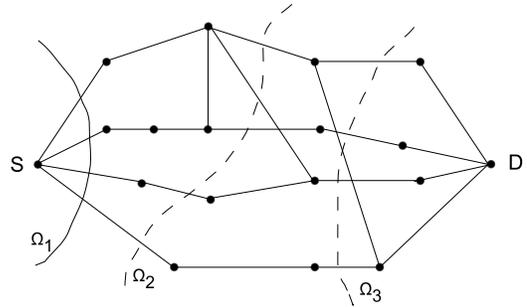}
\caption{Cuts in a network. Here, the min-cut is $\Omega_1$.
\label{fig:mincut}}
\end{figure}

An example of the dominating min-cut is shown in
Fig.~\ref{fig:mincut}.

\subsubsection{Amplify and Forward Protocols\label{sec:af}}

By an AF protocol $\wp$, we will mean a protocol $\wp$ in which each node in
the network operates in an amplify-and-forward fashion.  Such
protocols induce a linear channel model between source and sink of
the form:
 \beq
    \bold{y} \ = \bold{ H(\wp) {x} + {w} } \ ,
    \label{eq:channel_model}
\eeq where $\bold{y} \in \mathbb{C}^m$ denotes the signal received
at the sink, $\bold{w}$ is the noise vector, $\bold{H(\wp)}$ is
the $(m \times n)$ induced channel matrix and $\bold{x} \in
\mathbb{C}^n$ is the vector transmitted by the source. We impose
the following energy constraint on the vector $\bold{x}$
transmitted by the source, \beqan \text{Tr} \ (\Sigma_x) \ := \
\text{Tr} \ (\mathbb{E}\{\bold{x} \bold{x}^{\dagger}\}) & \leq & n
\rho, \eeqan where $\text{Tr}$ denotes the trace operator. We will
assume a symmetric energy constraint at the relays as well as the
source. Assuming the noise power spectral density to be equal to
$1$, $\rho$ corresponds to the SNR for any individual link. We
consider both half and full-duplex operation at the relay nodes.

Our attention here will be restricted to amplify-and-forward (AF)
protocols since as we shall see, this class of protocols can often
achieve the DMT of a network. More specifically, our protocol will
require the links in the network to operate according to a
schedule which determines the time slots during which a node
listens as well as the time slots during which it transmits. When
we say that a node listens, we will mean that the node stores the
corresponding received signal in its buffer. When a node does
transmit, the transmitted signal is simply a scaled version of the
most recent received signal contained in its buffer, with the
scaling constant chosen to meet a transmit power constraint.
\footnote{More sophisticated linear processing techniques would
include matrix transformations of the incoming signal, but turn
out to be not needed here.} In particular, nodes in the network
are not required to decode and then re-encode. It turns
out~\cite{AzaGamSch} that the value of the scaling constant does
not affect the DMT of the network operating under the specific AF
protocol.  Without loss of accuracy therefore, we will assume that
this constant is equal to $1$. It follows that, for any given
network, we only need specify the schedule to completely specify
the protocol. This will create a virtual MIMO channel of the form
${\bf y}\ = \ {\bf H}{\bf x} \ + \ {\bf w}$ where ${\bf H}$ is the
effective transfer matrix and ${\bf w}$ is the noise vector, which
is in general colored.

In following subsections of this section, we will develop
techniques to handle colored noise as well as establish results on
the DMT of some elementary network connections. We will also
establish lower bounds on the DMT of lower triangular matrices,
which will be useful later in computing the DMT of certain
protocols. We will also establish the maximum multiplexing gain
for channel matrices possessing certain structure.

\subsection{White in the Scale of Interest\label{sec:noise}}

In this section, we provide two results that will be extensively
used in all future sections: Theorem~\ref{thm:noise_white}, which
states that noise, even though correlated, can be treated as white
in the scale of interest and Lemma~\ref{lem:signal_white}, which
proves that i.i.d. gaussian inputs are sufficient to attain the
outage exponent of any channel of the form ${\bf y}= {\bf Hx}+
{\bf w}$.

If $\bold{h}$ is a Rayleigh random variable, then it is very easy
to see that, for any given $\epsilon$ and $\rho$, \beqan Pr \{
|\bold{h}|^2
> \rho^{\epsilon} \} \leq \exp(-\rho^{\epsilon}). \eeqan

Interestingly, a similar statement holds even when we replace $h$
by a polynomial in several Rayleigh random variables.

\blem \label{lem:tail_probability} Let $\{
\bold{h}_1,\bold{h}_2,...,\bold{h}_M \}$ be a collection of i.i.d.
Rayleigh random variables. Let $f \in \mathbb{C}[X_1,X_2,...,X_M]$
be a polynomial in the variables $X_i$ without a constant term.
Then there exists $A>0, B>0, d>0, \delta >0$ such that \beqan Pr
\{ |f(\bold{h}_1,\bold{h}_2,...,\bold{h}_M)|^2
> k \} & \leq & A \exp(-Bk^{\frac{1}{d}}) , \forall \ k \geq
\delta, \eeqan where the constants $A, B, d, \delta$ are
independent of k. \elem

\bpf See Appendix~\ref{app:tail_probability} \epf

We are now ready to establish that if the noise covariance matrix
has a certain structure, then it can be considered as white noise
for the purpose of DMT computation.

\bthm \label{thm:noise_white} Consider a channel of the form
$\bold{y} = \bold{Hx} + \bold{z}$. Let
$\bold{h}_1,\bold{h}_2,...,\bold{h}_L$ be $L$ i.i.d., Rayleigh
random variables. Let $\bold{G}_i,i=1,2,..,M$ be $N \times N$
matrices in which each entry is a polynomial function of the
random variables $\bold{h}_1,\bold{h}_2,...,\bold{h}_L$. Let
$\bold{z} = \bold{z}_0 + \sum_{i=1}^{M} \bold{G}_i\bold{z}_i$ be
the noise vector for a channel of the form ${\bf y = Hx + w}$. Let
$\{\bold{z}_i\}$ be independent $\mathbb{C}\mathcal{N}
(\underline{0},I)$ random vectors. Let the random matrix
$\bold{H}$ be a function of the random variables $\bold{h_i}$.
Then the noise vector $\bold{z}$ is white in the scale of
interest, i.e., the DMT of the channel ${\bf y = Hx + z}$ is the
same as the DMT of the channel ${\bf y = Hx + w}$ with ${\bf w}$
being a $\mathbb{C}\mathcal{N} (\underline{0},I)$ random vector.
\ethm \vspace{0.1in}

 \bpf See Appendix~\ref{app:noise_white}.
\epf \vspace{0.1in}
 \blem
\label{lem:signal_white} \cite{ZheTse} For any channel that is of
the form $\bold{y} = \bold{Hx} + \bold{w}$ with ${\bf w}$ being
white gaussian noise, i.i.d. gaussian inputs are sufficient to
attain the best possible outage exponent of the channel.\elem

\bpf  While a complete proof is available in \cite{ZheTse}, we
provide a sketch of the same proof for the sake of completeness.
The outage probability is given by, \beqan P_{\text{out}}(R) & = &
\inf_{\Sigma_x: \ \text{Tr} \ (\Sigma_x) \leq \mathbb{P}}
\Pr\{I(\bold{x};\bold{y} \mid \bold{H} = H) < R\} \eeqan

The mutual information is a function of the channel realization
and the distribution of the input. Nevertheless, without loss of
optimality, the distribution can be chosen to be gaussian, leading
to  \beqan P_{\text{out}}(R) & = & \inf_{\Sigma_x: \ \text{Tr} \
(\Sigma_x) \leq \mathbb{P}}
\Pr\{\log\det(I+\rho \bold{H}\Sigma_x \bold{H}^\dagger) < R\}. \\
\eeqan By bounding the eigenvalues of $\Sigma_x$, the outage
probability can be bounded below and above as,
\beqan \lefteqn{\Pr\{\log\det(I+\frac{\rho}{m}\bold{H} \bold{H}^\dagger) \leq R\}} \\
  & \geq & P_{\text{out}}(R) \geq \Pr\{\log\det(I+ \rho{\bold{H}}{\bold{H}}^\dagger) \leq R\}. \eeqan
As $\rho$ $\rightarrow$ $\infty$, it can be shown that the two
bounds converge so that we get Equation (9) in \cite{ZheTse}),
\beqa P_{\text{out}}(R) & \doteq & P(\log\det(I+\rho
\bold{H}\bold{H}^\dagger)<R). \eeqa \epf

The noise that we deal with in this paper will always satisfy the
conditions in Theorem~\ref{thm:noise_white}.  Hence we will make
the two assumptions appearing below throughout the paper: \bit
\item the transmitted signal has an i.i.d. gaussian distribution
\item the noise is white in the scale of interest. \eit

\subsection{DMT of Elementary Network Connections\label{sec:elementary_nw}}

\subsubsection{Parallel Network \label{sec:parallel} }

\begin{figure}[h!]
\centering
\includegraphics[height=50mm]{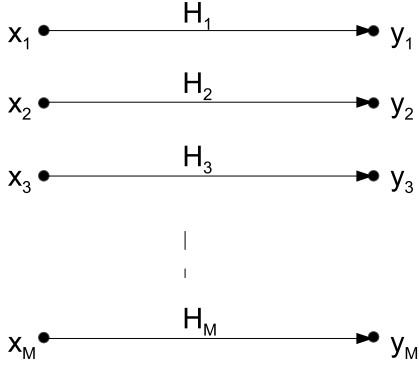}
\caption{The Parallel channel with M
sub-channels\label{fig:parallel_network}}
\end{figure}

The lemma below presents an expression for the DMT of a parallel channel in terms of the DMT of the individual links.

\blem \label{lem:parallel_channel} Consider a parallel channel
with $M$ links, with the $i$th link having representation ${\bf
y_i} = \bold{H_i}{\bf x_i} + \bold{w_i}$, and let $d_i(\cdot)$
denote the corresponding DMT.  Then the DMT of the overall
parallel channel is given by \beq d(r) =
\inf_{(r_1,r_2,\cdots,r_M): \ \sum_{i=1}^{M} r_i = r} \
\sum_{i=1}^{M} {d_i(r_i)} . \label{eq:parallel_dmt} \eeq \elem
\vspace{0.1in}

\bpf See Appendix~\ref{app:parallel_channel}. \epf \vspace{0.1in}

 The following lower and upper bounds on the outage exponent are
immediate from \eqref{eq:parallel_dmt}:
\beqa d(r) & \leq &  \sum_{i=1}^{M} {d_i\left(\frac{r}{K}\right)} \label{eq:dmt_parallel_upper} \\
d(r) & \geq &  \sum_{i=1}^{M} {d_i(r)}
\label{eq:dmt_parallel_lower} \eeqa

To determine the DMT of the parallel channel when all component channels are identically distributed with a DMT
that is a convex function of the rate, we will make use of the following Lemma from the theory of majorization
\cite{MarOlk}:

\blem \label{lem:majorization} \cite{MarOlk}  If $f(.)$ is a symmetric function in variables $r_1, r_2,\ldots,
r_N$ and is convex in each of the variables $r_i, i=1, 2,\ldots, N$, then, \beq \inf_{(r_1,r_2,\cdots,r_N): \
\sum_{i=1}^{N} r_i = r} \ f(r_1,r_2,...,r_N) = f \left ( \frac{r}{N},\frac{r}{N},...,\frac{r}{N} \right ) \eeq
\elem

\vspace{0.1in} The corollary below follows as a result. \bcor
\label{cor:parallel_identical} The DMT of a parallel channel with
all the individual channels being identical and having a convex
DMT is given by: \beqa d(r) & = & Md_1 \left ( \frac{r}{M} \right
). \label{eq:dmt_parallel} \eeqa \ecor \vspace{0.1in} The result
in Corollary~\ref{cor:parallel_identical} was also obtained in
\cite{YanBelMimoAf}.

\vspace{0.1in}
\subsubsection{Parallel Channel with Repeated Coefficients\label{sec:parallel_dependent}}

\blem \label{lem:parallel_dependent} Consider a parallel channel
with $M$ links and repeated channel matrices. More precisely, let
there be $N$ distinct channel matrices
$H^{(1)},H^{(2)},...,H^{(N)}$, with $H^{(i)}$ repeating in $n_i$
sub-channels, such that $\sum_{i=1}^{N}n_i = M$.

Then the DMT of such a parallel channel is given by, \beq d(r) =
\inf_{(r_1,r_2,\cdots,r_M): \ \sum_{i=1}^{N} \ n_i r_i = r} \
\sum_{i=1}^{N} {d_i(r_i)} . \label{eq:parallel_repeated_coeffs}
\eeq \elem

\vspace{0.2in}

\bpf The proof is along the lines of the proof of Lemma~\ref{lem:parallel_channel}, and is given in
Appendix~\ref{app:parallel_correlated}. \epf

\subsection{Maximum Multiplexing gain}

In this section, we derive the maximum multiplexing gain (MMG) of
a MIMO channel matrix with each entry of the matrix being a
polynomial function of certain Rayleigh random variables. We begin
by deriving certain properties of polynomial functions of gaussian
random variables and we will later use these characteristics to
obtain the MMG.

\blem \label{lem:poly_finite_regions} Let $p \in \mathbb{R}[X]$ be
any non-constant polynomial, and let its degree be $d$.   Consider
the set ${\cal R}$ of all $ x \in \mathbb{R}$ over which the
following two conditions are satisfied: \beq |p(x)| \leq k,
\label{eq:poly_local_condition_1}\eeq \beq |p^{\prime}(x)| \geq m
. \label{eq:poly_local_condition_2}\eeq

This subset  ${\cal R}$ of $\mathbb{R}$ can be expressed as the
union \beq {\cal R} \ = \ \cup_{i=1}^L R_i \eeq of disjoint
intervals $R_i \ = \ [a_i, b_i]$. Furthermore, $L \ \leq \ 2d$.
\elem

\vspace{0.1in}

\bpf See Appendix~\ref{app:poly_finite_regions} \epf
\vspace{0.1in} \blem \label{lem:poly_probability} Let $\{
\bold{x}_1,\bold{x}_2,...,\bold{x}_N \}$ be a collection of
independent gaussian random variables. Let $f \in
\mathbb{R}[X_1,X_2,...,X_N]$ be a polynomial in the variables
$X_i$. Then there exists constants $A>0,
 d>0, K>0$ such that \beqan \Pr \{ |f(\bold{x}_1,\bold{x}_2,...,\bold{x}_N)| < \delta \} &
\leq & A {\delta}^{\frac{1}{d}}, \ \ \forall \ 0 \leq \delta < K,
\eeqan where the constants $A, d, K$ depend only on $f$ and not on
$\delta$. \elem \vspace{0.1in}

\bpf See Appendix~\ref{app:poly_probability} \epf \vspace{0.1in}

We will proceed to utilize this lemma to obtain the MMG.

\bdefn Given a random matrix $\bold{H}$, which is a function of
random variables $\bold{h}_1,\bold{h}_2,\ldots,\bold{h}_N$, we
define the structural rank of $\bold{H}$ as the maximum rank
attained by $\bold{H}$, where the maximum is computed over all
possible realizations of the  $\{ \bold{h}_i \}$. We denote the
structural rank of a random matrix $\bold{H}$ by
$\mathrm{\mathbb{R}ank}(\bold{H})$. \edefn \vspace{0.1in}

\bthm \label{thm:DOF_rank} Consider a channel of the form
$\bold{y} = \bold{Hx + w}$, where $\bold{H} \in \mathbb{C}^{N
\times N}$ is a random matrix, and $\bold{x},\bold{y},\bold{w}$
are $N$-length column vectors representing the transmitted signal,
received signal and the noise vector respectively, with the noise
being white in the scale of interest. If the entries of ${\bf H}$
are polynomial functions of certain underlying Rayleigh random
variables, then the maximum multiplexing gain $D$ of the channel
is given by, \beqn D = \mathrm{\mathbb{R}}\text{ank} (\bold{H}).
\eeqn \ethm

\vspace{0.1in} \bpf We will prove that the MMG of the channel is
equal to the structural rank $\mathrm{\mathbb{R}ank}(\bold{H})=:m$
of $\bold{H}$. Clearly, for any given ${\bf H}$, the MMG is
upper-bounded by the rank of ${\bf H}$, which is lesser than $m$.
Therefore the upper-bound of $m$ on the MMG is clear. Next, we
will show that a MMG of $m$ is achievable, i.e., for any $\delta >
0$, a multiplexing gain of $(m - \delta)$ yields a non-zero
diversity gain.

Consider transmission at a multiplexing gain of $r = m - \delta$.
Since ${\bf H}$ is of structural rank $m$, there is a $m \times m$
sub-matrix $H_m$ of structural rank $m$. Then ${\bf H}_m {\bf
H}_m^{\dagger}$ is a principal sub-matrix of ${\bf H} {\bf
H}^{\dagger}$. Using the inclusion principle (\emph{Theorem}
$4.3.15$ in \cite{HorJoh}) and the fact that only $m$ eigenvalues
of ${\bf H}$ are non-zero, we obtain that, \beqan \log \det (I +
\rho {\bf H}{\bf H}^{\dagger}) \geq \log \det (I + \rho {\bf H}_m
{\bf H}_m^{\dagger}). \eeqan

Therefore, we get the outage exponent for rate $r = m-\delta$ as
\beqan \rho^{-d_{out}(r)} & = & \Pr\{ \log\det(I + \rho {{\bf H}}{{\bf H}^{\dagger}}) < (r)\log\rho \} \\
& \leq & \Pr\{ \log\det(I + \rho {{\bf H}_m}{{\bf H}_m^{\dagger}}) < (r)\log\rho \} \\
\rho^{-d_{out}(m-\delta)} & \leq & \Pr\{ \det(I + \rho {{\bf H}_m}{{\bf H}_m^{\dagger}}) < \rho^{(m-\delta)} \} \\
& \leq & \Pr\{ \det(\rho {{\bf H}_m}{{\bf H}_m^{\dagger}}) < \rho^{(m-\delta)} \} \\
& = & \Pr\{ \det({{\bf H}_m}{{\bf H}_m^{\dagger}}) < \rho^{-\delta} \} \\
& = & \Pr\{ |\det({\bf H}_m)|^2 < \rho^{-\delta} \}. \eeqan

Let the random matrix ${\bf H}$, and thereby its sub-matrix ${\bf
H}_m$, be a function of the Rayleigh random variables $\bold{h}_1,
\bold{h}_2, \ldots, \bold{h}_N$. Let us denote the real and
imaginary parts of this collection of Rayleigh random variables by
$\bold{x}_1,\bold{x}_2,\ldots,\bold{x}_N$, where $N = 2M$. Now
$\bold{x}_i$ are i.i.d. gaussian random variables, i.e., they are
distributed as $\mathcal{N}(0,1)$. Then $|\det(\bold{H}_m)|^2$ is
a non-zero real polynomial
$p(\bold{x}_1,\bold{x}_2,\ldots,\bold{x}_N)$ in $\bold{x}_i$.
Since $p =|\det(\bold{H}_m)|^2$ is positive,
$|p(\bold{x}_1,\bold{x}_2,\ldots,\bold{x}_N)| =
p(\bold{x}_1,\bold{x}_2,\ldots,\bold{x}_N)$.

We can now use Lemma~\ref{lem:poly_probability} to obtain that
\beqa \Pr\{ |\det(\bold{H}_m)|^2 < \rho^{-\delta} \}
& = & \Pr\{ |p(\bold{x}_1, \bold{x}_2, \ldots, \bold{x}_N)| < \rho^{-\delta} \} \nonumber \\
& \leq & A \rho^{-\delta / d}, \label{eq:MMG_interim} \eeqa for
some positive constants $A,d,K$ with $\rho^{-\delta} < K$. Let
$\rho_0^{-\delta} = K$. Then we can see that
\eqref{eq:MMG_interim} is valid for all $\rho > \rho_0$.

This leads to, \beqan \rho^{-d_{out}(m - \delta)} & \leq & A \rho^{-\delta/d}, \ \ \forall \rho > \rho_0 \\
\Rightarrow \rho^{-d_{out}(m - \delta)} & \dot \leq & \rho^{-\delta/d} \\
\Rightarrow  d_{out}(m - \delta) & \geq & {\delta}/{d} \\
& > & 0. \eeqan

Thus a MMG of $m$ is achievable and this concludes the proof.

\epf

\subsection{A Lower Bound on the DMT of Block-Lower-Triangular Matrices}

In this section, we give a lower bound on the DMT of ``block-lower-triangular''(blt) matrices, that are defined below. In many
situations, the matrices induced by AF protocols in a ss-ss
network will turn out to posses block-lower-triangular structure.

\bdefn \label{defn:Block_Lower_Triangular} Consider a set of $N_i \times N_j$ matrices $A_{ij},j=1,2,...,N, i
\geq j$. Let $A$ be the blt matrix comprised of the block matrices $A_{ij}$ in the
$(i,j)$th position and zeros elsewhere, i.e., \beqan A & = & \left[\begin{array}{cccc}
    A_{11} & 0 & \ldots & 0\\
    A_{21} & A_{22} & \ldots & 0\\
    \vdots & & \ddots &\vdots\\
    A_{N1} & A_{N2} & \ldots & A_{NN} \\
    \end{array}\right].
\eeqan We define the $l$-th sub-diagonal matrix, $A^{(\ell)}$ of such a blt matrix $A$ as the matrix comprising
only of the entries $A_{{\ell}1},A_{({\ell}+1)2},...,A_{({\ell}+N-1)N}$ with zeros everywhere else i.e.,
\beqan (A^{(\ell)})_{ij} & = & \left\{ \begin{array}{c} A_{ij} \ \text{ if } \ i-j=l-1, \\
                                         0_{N_i \times N_j}     \ \text{ otherwise. } \end{array}
                                         \right. \eeqan
The {\em last} sub-diagonal matrix of $A$ is defined as the
sub-diagonal matrix $A^{(\ell)}$ of $A$, where $\ell$ is the
largest integer for which $A^{(\ell)}$ is non-zero. Thus, for
example, the matrix whose only nonzero terms are the diagonal
entries of $A$ corresponds to $A^{(\ell)}$ with $\ell=0$ and the
matrix whose only nonzero entry is $A_{N1}$ corresponds to
$A^{(\ell)}$ with $\ell=(N-1)$.  \edefn

The theorem below establishes lower bounds on the DMT of channel matrices which have a blt structure.

\bthm \label{thm:main_theorem} Consider a random blt matrix $\bold{H}$ having component matrices $\bold{H}_{ij}$ of size
$N_i \times N_j$. Let $M := \sum_{i=1}^{N} N_i$ be the size of the square matrix $\bold{H}$.

Let $\bold{H}^{(0)}$ be the diagonal part of the matrix $\bold{H}$ and $\bold{H}^{(\ell)}$ denote the last
sub-diagonal matrix of $\bold{H}$, as given by Definition~\ref{defn:Block_Lower_Triangular}. Then, \ben \item $d_H(r)
\geq d_{H^{(0)}}(r)$. \item $d_H(r) \geq d_{H^{(\ell)}}(r)$.

\item In addition, if the entries of $H^{(\ell)}$ are independent
of the entries in $H^{(0)}$, then $d_H(r) \geq d_{H^{(0)}}(r) + d_{H^{(\ell)}}(r)$. \een \ethm

\begin{figure*}
  \centering
  \subfigure[The i-f diagram]{\label{fig:if_block_lt}\includegraphics[width=40mm]{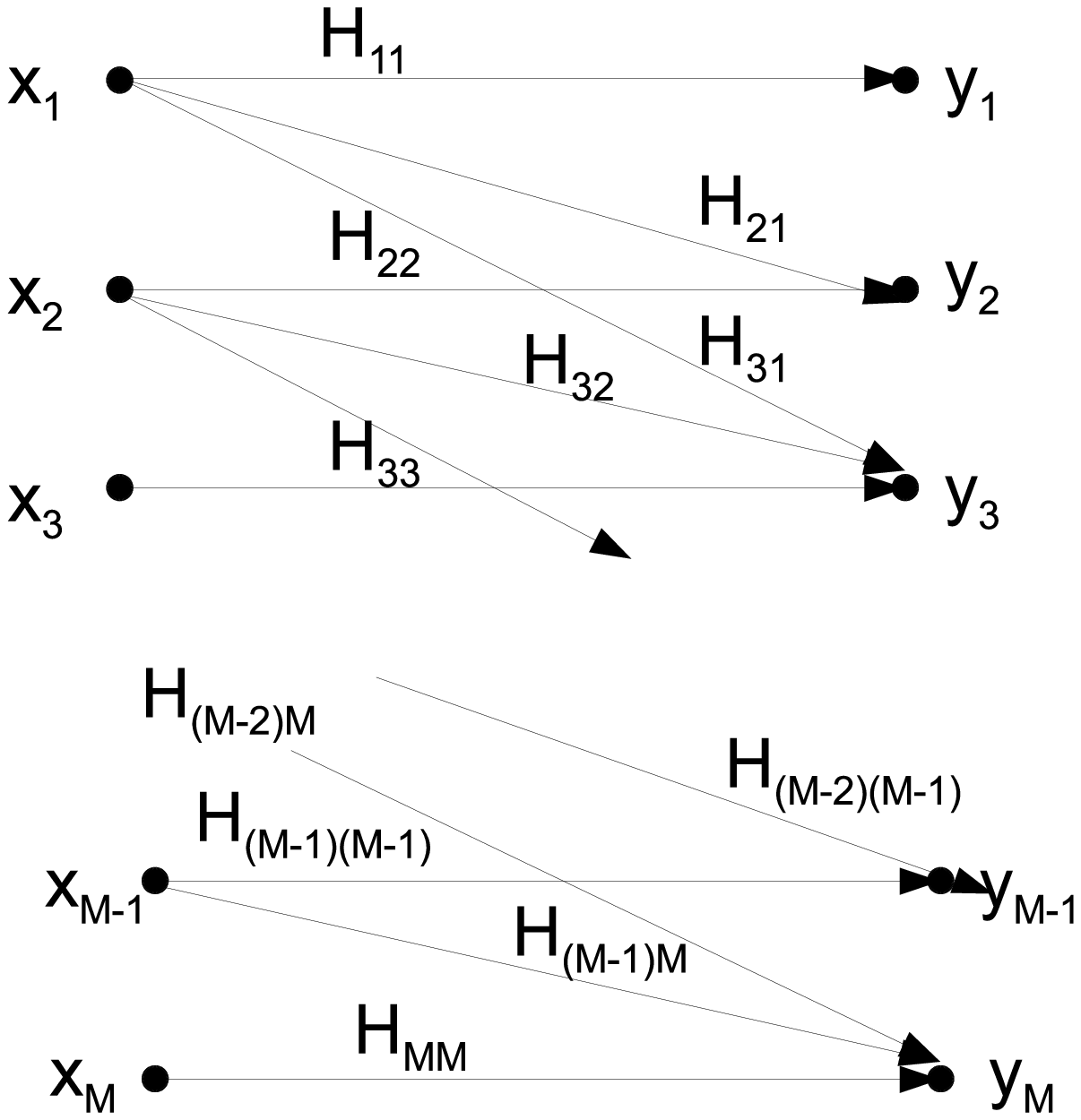}}
  \hspace*{0.3in} \subfigure[Showing only the direct links]{\label{fig:if_block_lt_direct}\includegraphics[width=40mm]{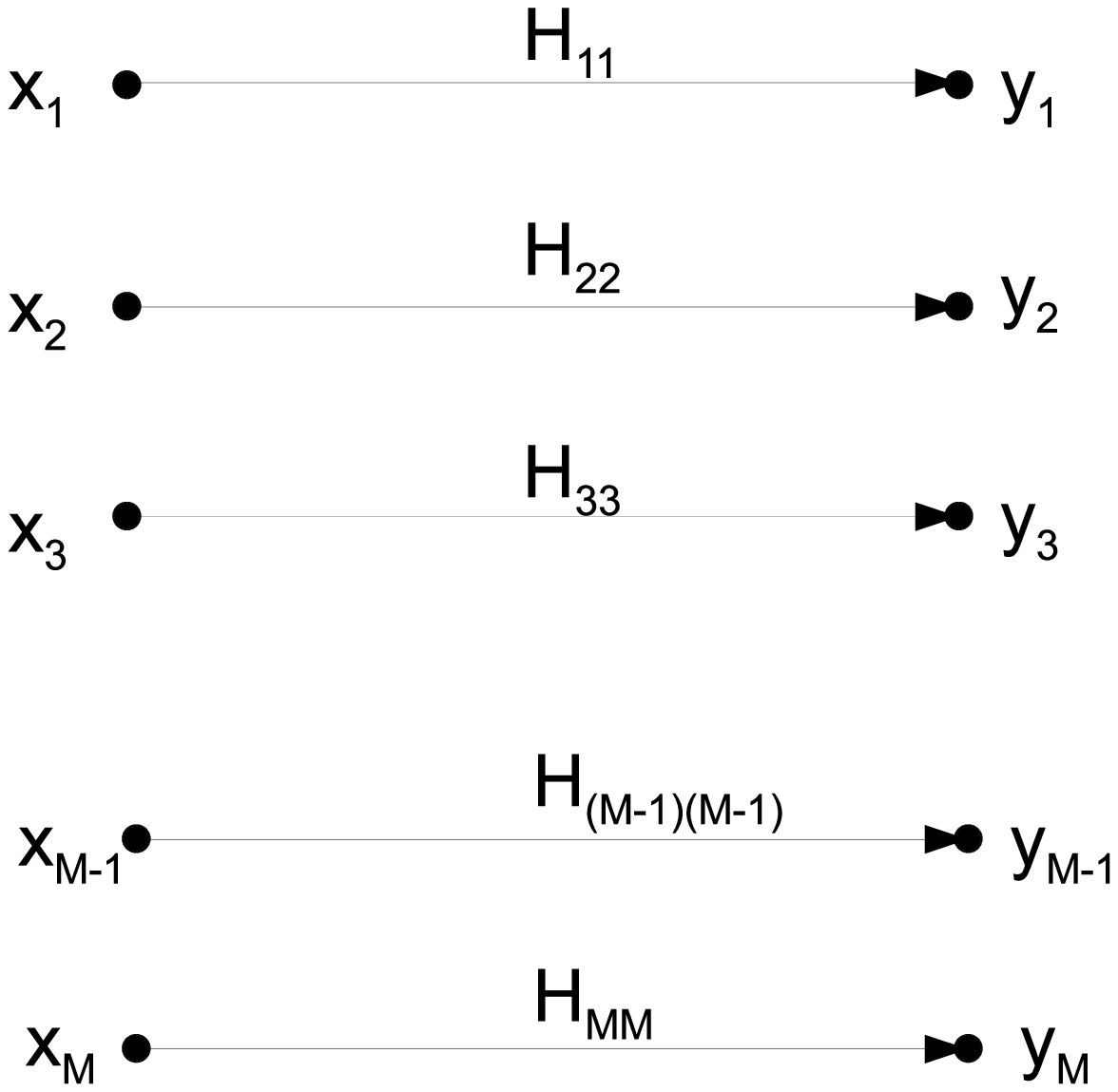}}
    \hspace*{0.3in} \subfigure[Showing only cross links.]{\label{fig:if_block_lt_cross}\includegraphics[width=40mm]{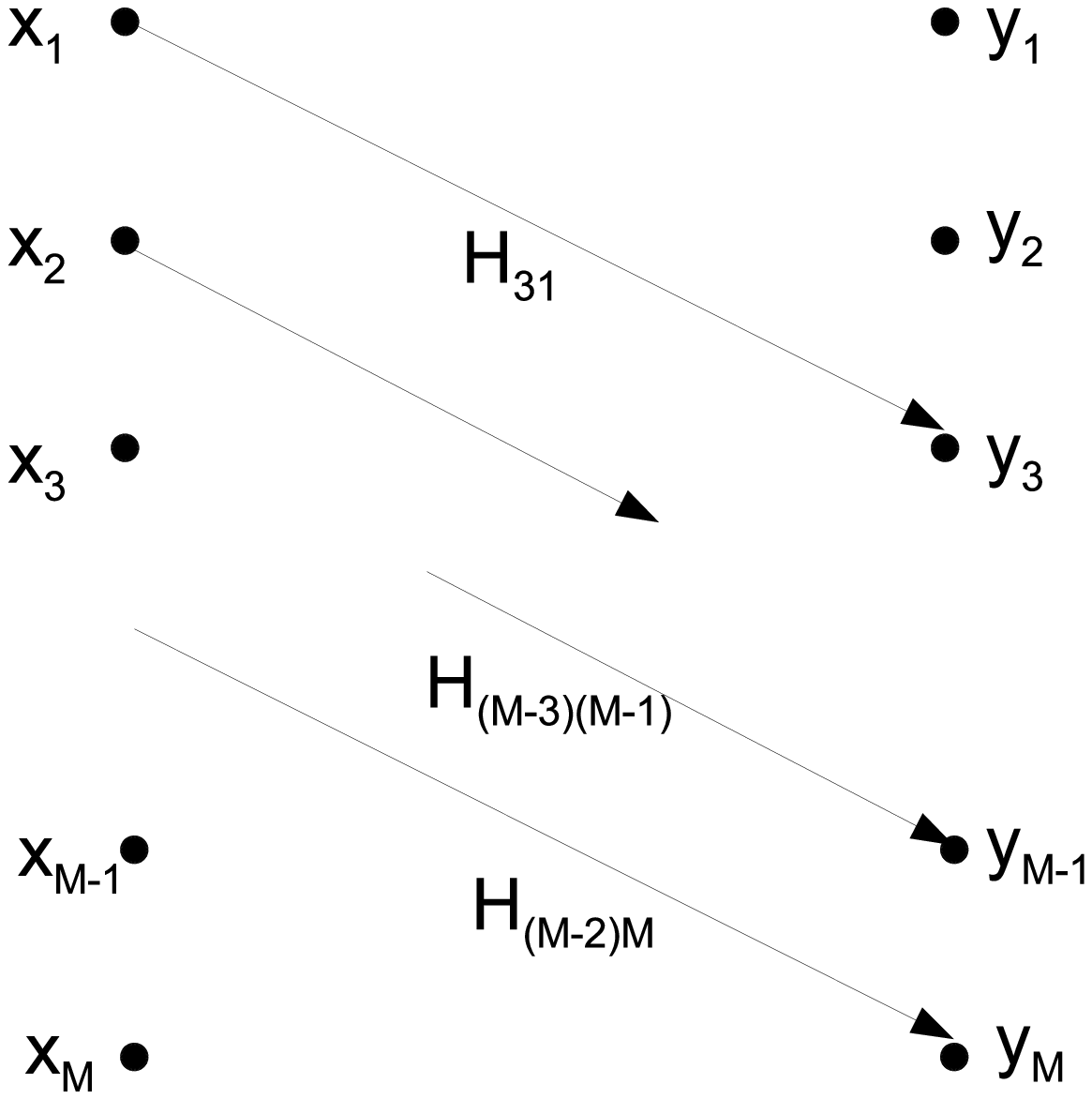}}
  \caption{The i-f diagram for the block lower triangular matrix and its decomposition.}
  \label{fig:if_block_lt_decomposition}
\end{figure*}

\bpf The channel is given by $\bold{y} = \bold{H}\bold{x} +
\bold{w}$. Since the noise is white in the scale of interest, by
Theorem~\ref{thm:noise_white}, the DMT of this channel is the same
as that of a channel with the noise distributed as
$\mathbb{CN}(0,I)$. Therefore, without loss of generality, we
assume that $\bold{w}$ is distributed as $\mathbb{CN}(0,I)$.

For any given matrix $H$, the outage probability
exponent\cite{ZheTse} is given by \beqan \rho^{-d_{H}(r)} & \doteq
& \inf_{\Sigma_x: \text{Tr} \ (\Sigma_x) \leq \mathbb{P}}
\Pr\{I(\bold{x};\bold{y}:\bold{H} = H) \leq r\log\rho\}.
\label{eq:outage_exponent} \eeqan To estimate this exponent, we
begin by identifying lower bounds on the mutual information. Note
that by Lemma~\ref{lem:signal_white}, for the purposes of
computing outage exponent, we may assume without loss of
optimality that the input $\bold{x}$ is distributed as
$\mathbb{C}\mathcal{N}(0,I)$. We will make this assumption.

Due to the fact that the last sub-diagonal matrix is given by the
${\ell}$-th sub-diagonal matrix, we have, \beqan \bold{y}_i =
\bold{H}_{ii} \bold{x}_i + \bold{H}_{i(i-1)} \bold{x}_{i-1} + ...
+ \bold{H}_{i(i-{\ell})} \bold{x}_{i-{\ell}} + \bold{w}_i. \eeqan

Starting with the mutual information term, we have,  \beq
I(\bold{x};\bold{y} |\bold{H}=H) = \sum_{i=1}^{M}
I(\bold{x}_i\bold{y} | \bold{H} = H, \bold{x}_1^{i-1}) . \eeq
Consider next, the following series of inequalities for all $i=1,
\ldots, N$:
\beqa \lefteqn{I(\bold{x}_i;\bold{y} | \bold{H} = H, \bold{x}_1^{i-1})} \nonumber \\
& \geq & I(\bold{x}_i;\bold{y}_i | \bold{H} = H, \bold{x}_1^{i-1}) \nonumber \\
& = & I(\bold{x}_i; \bold{H}_{ii} \bold{x}_i + \bold{H}_{i(i-1)} \bold{x}_{i-1}  \nonumber \\
&  & + \ldots + \bold{H}_{i(i-{\ell})} \bold{x}_{i-{\ell}} +
\bold{w}_i | \bold{H} = H, \bold{x}_1^{i-1}) \nonumber \eeqa
\beqa & = & I(\bold{x}_i; H_{ii} \bold{x}_i + H_{i(i-1)} \bold{x}_{i-1} \nonumber \\
&  & + \ldots + H_{i(i-{\ell})} \bold{x}_{i-{\ell}} +  \bold{w}_i | \bold{H} = H, \bold{x}_1^{i-1}) \nonumber \\
& = & I(\bold{x}_i; H_{ii} \bold{x}_i + H_{i(i-1)} \bold{x}_{i-1} \nonumber \\
&  & + \ldots + H_{i(i-{\ell})} \bold{x}_{i-{\ell}}+  \bold{w}_i | \bold{x}_1^{i-1}) \nonumber \\
& = & I(\bold{x}_i; H_{ii} \bold{x}_i  +  \bold{w}_i | \bold{x}_1^{i-1}) \nonumber \\
& = & I(\bold{x}_i; H_{ii} \bold{x}_i  +  \bold{w}_i ). \nonumber
\eeqa The last step follows since $\{ \bold{ x_i } \}$ are
independent under the assumed $\mathbb{C}\mathcal{N}(0,I)$
distribution. We thus have, \beqa I(\bold{x};\bold{y} |\bold{H}=H)
& \geq &  \sum_{i=1}^{M} I(\bold{x}_i; H_{ii} \bold{x}_i  +  \bold{w}_i) \nonumber \\
& \geq & I(\bold{x};  \bold{H^{(0)}x + w} | \bold{H^{(0)}}=H^{(0)}). \nonumber \\
\label{eq:main_diagonal_bound} \eeqa

In the above, as is customary, whenever a variable with a negative index is encountered, it should be
interpreted as if the variable were not present.  From \eqref{eq:main_diagonal_bound}, it follows that \beqa \rho^{-d_H(r)} & \dot = &
\Pr\{I(\bold{x};\bold{y}|\bold{H} = H) \leq r\log\rho\} \\
& \leq & \Pr\{ I(\bold{x};  \bold{H}^{(0)}\bold{x} + \bold{w} |\bold{H}^{(0)} = H^{(0)})  \leq r\log\rho \}\nonumber \\
& \dot = & \rho^{-d_{H^{(0)}}(r)}, \nonumber \eeqa i.e., \beqa d_H(r) & \geq & d_{H^{(0)}}(r). \eeqa

In the ``information-flow diagrams'' appearing in Fig.~\ref{fig:if_block_lt_decomposition}, the lower bounding
of the mutual information by replacing the matrix $H$ by the diagonal $H^{(0)}$ can be seen to correspond to a
pruning of the graph shown in Fig.~\ref{fig:if_block_lt} resulting in the figure in
Fig.~\ref{fig:if_block_lt_direct}.

Similarly, we have a second set of inequalities which correspond effectively to replacing the matrix $H$ by the
last sub-diagonal matrix $H^{(\ell)}$. This corresponds to the pruned graph appearing in
Fig.~\ref{fig:if_block_lt_cross}.
\beqa \lefteqn{I(\bold{x}_{i-{\ell}};\bold{y} | \bold{H} = H, \lefteqn{\bold{x}_{i-{\ell}+1}^N)}} \nonumber \\
& \geq & I(\bold{x}_{i-{\ell}};\bold{y}_{i} | \bold{H} = H, \bold{x}_{i-{\ell}+1}^N) \nonumber \\
& = & I(\bold{x}_{i-{\ell}}; \bold{H}_{ii} \bold{x}_i + \bold{H}_{i(i-1)} \bold{x}_{i-1} + \ldots + \nonumber \\
&  & \bold{H}_{i(i-{\ell})} \bold{x}_{i-{\ell}} +  \bold{w}_{i} |
\bold{H} = H, \bold{x}_{i-{\ell}+1}^N) \nonumber \eeqa
\beqa & = & I(\bold{x}_{i-{\ell}}; H_{ii} \bold{x}_i + H_{i(i-1)} \bold{x}_{i-1} + \ldots + \nonumber \\
& & H_{i(i-{\ell})} \bold{x}_{i-{\ell}} +  \bold{w}_{i}  | \bold{H} = H, \bold{x}_{i-{\ell}+1}^N ) \nonumber \\
& = & I(\bold{x}_{i-{\ell}}; H_{ii} \bold{ x}_i + H_{i(i-1)}
\bold{ x}_{i-1} + \ldots + \nonumber \\
 & & H_{i(i-{\ell})} \bold{ x}_{i-{\ell}} + \bold{w}_{i}  | \bold{x}_{i-{\ell}+1}^N) \nonumber \\
& = & I(\bold{x}_{i-{\ell}}; H_{i(i-{\ell})} \bold{ x}_{i-{\ell}} +  \bold{w}_{i}  | \bold{x}_{i-{\ell}+1}^N) \nonumber \\
& = & I(\bold{x}_{i-{\ell}}; H_{i(i-{\ell})} \bold{ x}_{i-{\ell}} +  \bold{w}_{i}) \nonumber \eeqa
\beqa \Rightarrow \lefteqn{I(\bold{x};\bold{y} |\bold{H}=H)} \nonumber \\
& = &  \sum_{i=N}^{1} I(\bold{x}_i;\bold{y} | \bold{H} = H, \bold{x}_{i+1}^N ) \nonumber \\
& \geq &  \sum_{i=N}^{l+1} I(\bold{x}_{i-{\ell}};\bold{y} | \bold{H} = H, \bold{x}_{i-{\ell}+1}^N ) \nonumber \\
& \geq &  \sum_{i=N}^{l+1} I(\bold{x}_{i-{\ell}}; H_{i(i-{\ell})} \bold{ x}_{i-{\ell}} +  \bold{w}_{i})\nonumber \\
& = & I(\bold{x}; \bold{H}^{(\ell)}x + \bold{w} |
\bold{H}^{(\ell)}=H^{(\ell)}). \label{eq:sub_diagonal_bound} \eeqa

Thus \beqa \rho^{-d_H(r)} & \dot = &  \Pr\{I(\bold{x};\bold{y}|\bold{H} = H) \leq r\log\rho\} \\
& \leq & \Pr\{ I(\bold{x};  \bold{H}^{(\ell)}\bold{x} + \bold{w} |\bold{H}^{(\ell)} = H^{(\ell)})  \leq r\log\rho \}\nonumber \\
& \dot = & \rho^{-d_{H^{(\ell)}}(r)} \nonumber \\
\Rightarrow \ d_H(r) & \geq & d_{H^{(\ell)}}(r). \eeqa

It follows therefore, from \eqref{eq:main_diagonal_bound} and \eqref{eq:sub_diagonal_bound} that \beqa
\lefteqn{I(\bold{x};\bold{y} |\bold{H}=H)} \nonumber \\
& \geq & \max(I(\bold{x}; \bold{H_dx + w} | \bold{H_d}=H_d), \nonumber \\
& & I(\bold{x}; \bold{H^{(\ell)}x + w} | \bold{H^{(\ell)}}=H^{(\ell)})). \label{eq:max_bound}\eeqa

This leads to
\beqa \rho^{-d_H(r)} & \dot = &  \Pr\{I(\bold{x};\bold{y}:\bold{H} = H) \leq r\log\rho\} \nonumber \\
& \leq & \Pr\{ \max \{I(\bold{x};  \bold{H}^{(0)}\bold{x}  +  \bold{w} | \bold{H}^{(0)} = H^{(0)}), \nonumber \\
& & I(\bold{x}; \bold{H}^{(\ell)}\bold{x} + \bold{w} | \bold{H}^{(\ell)}=H^{(\ell)})\} \leq r\log\rho \}\nonumber \\
& = & \Pr\{ I(\bold{x};  \bold{H}^{(0)}\bold{x} + \bold{w} |
\bold{H}^{(0)}=H^{(0)})\leq r\log\rho, \nonumber \\
& & I(\bold{x}; \bold{H}^{(\ell)}\bold{x} + \bold{w} |
\bold{H}^{(\ell)}=H^{(\ell)})) \leq r\log\rho \}\nonumber \eeqa
\beqa & = & \Pr\{ I(\bold{x};  \bold{H}^{(0)}\bold{x} + \bold{w} | \bold{H}^{(0)}=H^{(0)})\leq r\log\rho \} \nonumber \\
& & . \ \Pr\{ I(\bold{x};  \bold{H}^{(\ell)}\bold{x} + \bold{w} |
\bold{H}^{(\ell)}=H^{(\ell)})\leq r\log\rho \} \nonumber \\
 & \dot = & \rho^{-d_{H^{(0)}}(r)} \rho^{-d_{H^{(\ell)}}(r)} \nonumber \\
\ & \dot = & \rho^{-d_{H^{(0)}}(r)+d_{H^{(\ell)}}(r)}  \nonumber \\
\Rightarrow \ d_H(r) & \geq & d_{H^{(0)}}(r)+d_{H^{(\ell)}}(r). \eeqa

where the first step comes about because of the independence of
the entries in $\bold{H^{(0)}}$ and $\bold{H^{(\ell)}}$, which is
indeed the case. \epf

\vspace{0.1in} \bnote \label{rem:Matrix_Inequality} The following
two matrix inequalities can be deduced from the proof of
Theorem~\ref{thm:main_theorem}, with $\bold{H}^{(0)}$ and
$\bold{H}^{(\ell)}$ defined as in the theorem: \beqa \det ( I  +
\rho \bold{H} \bold{H}^\dagger ) & \geq &  \det ( I  + \rho
\bold{H}^{(0)} \bold{H}^{(0)\dagger} ) \\
\det ( I  + \rho \bold{H} \bold{H}^\dagger ) & \geq &  \det ( I  +
\rho \bold{H}^{(\ell)} \bold{H}^{(\ell)\dagger} ). \eeqa \enote

\vspace{0.1in} \bnote Although the result is derived for lower
triangular matrices, it also applies in a slightly more general
setting. Consider a band matrix of the form given below, \beqan
\bold{H} & = & \left[\begin{array}{cccccc}
    \ast &  & \ast & & & \\
         & \ast &     & \ast & & \\
    \ast &  &  \ast    & & \ast& \\
         & \ast &      & \ast & & \ast \\
         &  & \ast & & \ast &  \\
         &  &  & \ast & & \ast \\
    \end{array}\right],
\eeqan

where there are bands of non-zero entries, denoted by sequence of
$\ast$. Let $\bold{H}_{ub}$ and $\bold{H}_{lb}$ denote matrices
derived from $\bold{H}$, constituting of only the uppermost band
and lowermost band respectively. They will be, respectively of the
form, \beqn \bold{H}_{ub} =  \left[\begin{array}{cccccc}
     &  & \ast & & & \\
         &  &     & \ast & & \\
     &  &      & & \ast& \\
         &  &      &  & & \ast \\
         &  &  & &  &  \\
         &  &  &  & &  \\
    \end{array}\right], \ \
\bold{H}_{lb}  =  \left[\begin{array}{cccccc}
     &  &  & & & \\
         &  &     &  & & \\
    \ast &  &     & & & \\
         & \ast &      & & &  \\
         &  & \ast & &  &  \\
         &  &  & \ast & &  \\
    \end{array}\right].
\eeqn

Without affecting the DMT, the matrix $\bold{H}$ can be
transformed to a blt matrix of larger size, by adding an
appropriate number of all-zero rows at the top and all-zero
columns to the right. Then the uppermost band of the $\bold{H}$
belongs the diagonal, and the lowermost band belongs to the last
sub-diagonal of the new matrix. By invoking
Theorem~\ref{thm:main_theorem} for the new matrix, we get \ben
\item $d_H(r) \geq d_{H_{ub}}(r)$. \item $d_H(r) \geq
d_{H_{\ell{b}}}(r)$. \een If the entries of $\bold{H}_{{\ell}b}$
and $\bold{H}_{ub}$ are independent of each other, then we further
have, \beqn d_H(r) \geq d_{H_{ub}}(r) + d_{H_{\ell{b}}}(r). \eeqn

\enote

\vspace{0.2in}

\subsubsection{Example Applications of the DMT Lower bound\label{sec:examples_main_thm}}

In this subsection, we derive lower bounds to the DMT of two-hop
networks under the operation of various existing AF protocols. One
lower bound proves a conjecture by Rao and Hassibi~\cite{RaoHas},
while a second is tighter than lower bounds known earlier. In the
remaining instances, although the results do not add to what is
already known, the derivations presented here are surprisingly
simple and provide some intuitive explanation as to how these
protocols achieve the DMT.

\vspace{0.05in}

\emph{Example 1: Single relay, NAF protocol}

\vspace{0.05in}

Consider the relay network in Fig.\ref{fig:one_relay}. Let
$\bold{g}_d$, $\bold{g}_1$, $\bold{h}_1$ denote the channel
coefficients along the links from source to the sink, source to
the relay and relay to the sink respectively. The induced channel
under the NAF protocol is given by,
\beqa \left[\begin{array}{c} \bold{y}_1 \\
              \bold{y}_2 \end{array} \right] & = &
    \left[ \begin{array}{cc} \bold{g}_d & 0\\
     \bold{g}_1\bold{h}_1  & \bold{g}_d \end{array} \right] \left[\begin{array}{c} \bold{x}_1 \\
                                                           \bold{x}_1 \end{array} \right] +
\left[ \begin{array}{c} \bold{w}_1 \\
    \bold{w}_2  \  +  \ h_1  \bold{v}
    \end{array} \right] \nonumber \\
    \label{eq:naf_model}
\eeqa

Since two time instants are used in order to obtain the equivalent
channel matrix, $\bold{H}$, we have a rate loss by a factor of 2,
and hence $d(r) = d_H(2r)$. It can be checked that the noise is
white in the scale of interest. Now it is sufficient to study the
DMT of the matrix $H$. Let
\beqan \bold{H}^{(0)} & = & \left[ \begin{array}{cc}  \bold{g}_d & 0\\
     0  & \bold{g}_d \end{array} \right], \text{    and}\\
      \bold{H}^{(\ell)} & = & \left[ \begin{array}{cc} 0 & 0\\
     \bold{g}_1\bold{h}_1  & 0 \end{array} \right]. \eeqan

The fading coefficients $\bold{g}_d$, $\bold{g}_1$, $\bold{h}_1$
are independent and therefore $\bold{H}^{(0)}$ is independent of
$\bold{H}^{(\ell)}$. Invoking Theorem~\ref{thm:main_theorem} we
obtain: \beqan d_{H}(r) & \geq & d_{H^{(0)}}(r) +
d_{H^{(\ell)}}(r) \eeqan The diversity gains $d_{H^{(0)}}(r)$ and
$d_{H^{(\ell)}}(r)$ are easily evaluated as,
\beqan d_{H^{(0)}}(r) & = & \left(1 - \frac{r}{2}\right)^{+} ,\\
d_{H^{(\ell)}}(r) & = & (1 - {r})^{+}.\\
\text{ This leads to } d_H(r) & \geq & \left(1 -
\frac{r}{2}\right)^{+} + (1 - {r})^{+} . \eeqan

This leads to the following estimate of the DMT $d(r)$ of the protocol:
\beqan  d(r) & = & d_H(2r)\\
\Rightarrow d(r) & \geq & (1 - {r})^{+} + (1 - 2r)^{+}. \eeqan

From \cite{AzaGamSch} we know that this bound is indeed tight.
\bnote For the case of NAF protocol used with $N$ relays, it can be shown that Theorem~\ref{thm:main_theorem}
can be used to obtain a lower bound on the DMT of NAF protocol as \beqa d(r) & \geq & (1 - {r})^{+} + N(1 -
2r)^{+}. \eeqa This lower bound is proved to be tight for the $N$-relay case as well in \cite{AzaGamSch}. \enote

\vspace{0.05in}

\emph{Example 2: Multiple relays, SAF}

\vspace{0.05in}

Consider the network in Fig.\ref{fig:classical_relay} with $N$
relays. We employ an M-slot AF protocol termed the Slotted
Amplify-and-Forward (SAF) protocol and introduced in
\cite{YanBelSaf}. We assume that the relays are isolated from each
others' transmissions (see \cite{YanBelSaf} for a description).
Each symbol transmitted by the source reaches the sink through the
direct link, as well as through precisely one relayed path. For
this relay-isolated case, the induced channel matrix for a
$M$-slot protocol is given by a $M \times M$ matrix ${\bf H}$,
with $\bold{g}_d$, the fading coefficient of the direct link,
appearing along the diagonal, and with $\bold{g}_1, \ldots,
\bold{g}_N$, the product coefficients on the different relay
paths, appearing in repeated cyclic fashion along the first
sub-diagonal. Let $M = kN + 1$ denote the slot length, for a
positive integer $k$.

For example, in the $M=5$, $N=2$, $k=2$ case, the induced channel matrix ${\bf H}$ is given by, \beqan \bold{H}
& = & \left[
\begin{array}{ccccc}
\bold{g}_d & 0 & 0 & 0 & 0 \\
\bold{g}_1 & \bold{g}_d & 0 & 0 & 0  \\
0 & \bold{g}_2 & \bold{g}_d & 0 & 0  \\
0 & 0 & \bold{g}_1 & \bold{g}_d & 0 \\
0 & 0 & 0 & \bold{g}_2 & \bold{g}_d  \\ \end{array} \right].
\eeqan

Since the channel is used for $M$ time slots, we have the relation
$d(r) = d_H(Mr)$ between the DMT of the protocol, $d(r)$, and the
DMT of the matrix $d_H(r)$.  We next proceed to find a lower bound
on the DMT of the matrix.  As before, we use $\bold{H}^{(0)} =
\bold{g}_dI$ to denote the diagonal matrix associated to
$\bold{H}$.  Similarly, let $\bold{H}^{(\ell)}$ denote the last
sub-diagonal matrix corresponding to $\bold{H}$.   This matrix
contains $\bold{g}_1,\ldots,\bold{g}_N$ repeated $k$ times
cyclically along the first sub-diagonal. By
Theorem~\ref{thm:main_theorem}, the DMT of $\bold{H}$ can be lower
bounded as, \beqan
d_H(r) & \geq & d_{H^{(0)}}(r) + d_{H^{(\ell)}}(r) \\
\Rightarrow d(r) & \geq & d_{H^{(0)}}(Mr) + d_{H^{(\ell)}}(Mr) . \eeqan

The DMT of the matrices $\bold{H^{(0)}}$ and $\bold{H^{(\ell)}}$ can be easily derived as, $ d_{H^{(0)}}(r) =
\left(1 - \frac{r}{M}\right)^{+} $ and $ d_{H^{(\ell)}}(r) = N\left(1 - \frac{r}{M-1}\right)^{+}$ leading to:

\beqan d(r) & \geq & (1-r)^{+} + N\left(1 - \frac{M}{M-1}r\right)^{+}. \eeqan

The right hand side is in fact shown to be equal to the DMT of the SAF protocol in \cite{YanBelSaf} under the
assumption that relays are isolated.

\vspace{0.05in} \emph{Example 3: Multiple-Antenna, Single-Relay, NAF protocol} \vspace{0.05in}

Consider a single-relay network with the source, the relay and
sink equipped with multiple antennas given by $n_s$, $n_r$ and
$n_d$ respectively.  We follow \cite{YanBelMimoAf} and assume
operation under the NAF protocol introduced in \cite{AzaGamSch}
for the single-antenna case. The channel matrix turns out to be
given by,
\beqa \bold{H} & = & \left[ \begin{array}{cc}  \bold{H}_d & 0 \\
     \bold{H}_r  & \bold{H}_d \end{array} \right], \label{eq:MIMO_NAF}\eeqa

where $\bold{H_d}$ is the $n_d \times n_s$ fading matrix between
source and the sink, $\bold{H_r}$ is the product fading matrix of
an $n_r \times n_s$ Rayleigh fading matrix between the source and
the relay and an $n_d \times n_r$ Rayleigh fading matrix between
relay and sink. Proceeding in the same manner as in \emph{Example
1}, we get \beqan d(r) & \geq & d_{H_d}(r) + d_{H_r}(2r). \eeqan
This lower bound appears as Theorem 1 in \cite{YanBelMimoAf}.

\vspace{0.15in}

\emph{Example 4: Multiple Antenna, Multiple relays, NAF protocol} \vspace{0.05in}
We consider a $N$-relay network with each node in the network having multiple antennas. Let $n_s$,$n_{i}$ and
$n_d$ denote the number of antennas with the source, $i$th relay and the destination respectively.

The NAF protocol was proposed in \cite{AzaGamSch} for the case of
$N$ relays, with all nodes possessing single antennas. The
protocol can be viewed as using the NAF protocol for each relay
separately (the protocol comprises of two slots, with the source
transmitting to the relay and destination in the first slot and
the relay and the source transmitting to the destination in the
second slot) and then cycling through all the relays. The same
protocol was used in the case of multiple antenna relays in
\cite{YanBelMimoAf}. However it is not clear that this is the
optimal thing to do if each relay has different number of
antennas. In that case, we might want to use the relay with more
antennas more frequently in order to get a better performance.

Therefore, in this example, we consider a generalization of the
NAF protocol for multiple antenna relays, where we cycle through
all the relays for unequal periods of time. Specifically, we use a
NAF protocol for relay $i$ for $m_i$ cycles. When we say we use a
NAF protocol for relay $i$, it means that the source will first
transmit to the relay during the first time instant and then in
the second time instant the source and the relay will transmit to
the destination. Thus a NAF protocol operated for a single relay
for one cycle will take up $2$ time instants. Let $M:= \sum m_i$.
Then the protocol operates for $2M$ time slots and the induced
channel matrix $\bold{H}$ of size $2M \times 2M$ between source
and destination will contain the direct link fading matrix
$\bold{H}_d$ repeated along the diagonal and the first
sub-diagonal will have the product matrix $\bold{H}_i$
corresponding to relay $i$ repeated for $2m_i$ times.

More precisely, the relay matrix $\bold{H}_i$ is the product of
the Rayleigh fading matrix $\bold{F}_i$ between the source and the
$i$th relay and the Rayleigh fading matrix $\bold{G}_i$ between
the $i$th relay and the destination. Then the DMT $d_i(r)$ of the
product matrix $\bold{H}_i$ can be computed using the Rayleigh
product channel DMT in \cite{YanBelNew}.

This induces an effective channel matrix between the source and
the destination, which will be of the form:
\beqa \bold{H} & = & \left[ \begin{array}{cc}  \bold{H}_D & 0\\
     \bold{H}_R  & \bold{H}_D \end{array} \right], \label{eq:gen_NAF}\eeqa

where $ \bold{H}_D = I_{T} \otimes \bold{H}_d$ ,  with $I_T$
denoting the identity matrix of size $T$, $\otimes$ denoting the
tensor product and $\bold{H}_{d}$ denoting the $n_s \times n_d$
fading matrix corresponding to the direct link between the source
and the destination. $\bold{H}_R$ is now a block diagonal matrix
with the matrix $\bold{H}_i$ appearing along the diagonal for
$m_i$ times. Now the DMT of the protocol is given by \beqn d(r) =
 d_H(2Mr) \geq d_{H^{(0)}}(2Mr) + d_{H^{(\ell)}}(2Mr) \eeqn by
Theorem~\ref{thm:main_theorem}. Since $\bold{H}^{(0)}$ contains
the diagonal element as $H_d$ repeated $2M$ times,
$d_{H^{(0)}}(2Mr) = d_{H_d}(r)$. Also $d_{H^{(\ell)}}(2Mr) =
d_{H_R}(2Mr)$.

The DMT of the matrix $H_R$ can be computed using the DMT of
parallel channel with repeated coefficients
(Lemma~\ref{lem:parallel_dependent}).  This gives us \beqan
d_{H_R}(Mr) & = & \inf_{(r_1,r_2,\cdots,r_N): \ \sum_{i=1}^{N} \
f_i r_i = r} \ \sum_{i=1}^{N} {d_i(r_i)}, \eeqan where $f_i =
\frac{m_i}{M}$.

Thus a lower bound on $d(r)$ can be computed as \beqn d(r) \geq
d_{H_d}(r) + \inf_{(r_1,r_2,\cdots,r_N): \ \sum_{i=1}^{N} \ f_i
r_i = 2r} \ \sum_{i=1}^{N} {d_i(r_i)}. \eeqn

Now since the activation durations $\{f_i\}$ for the relays are
arbitrary, we can optimize the DMT over all possible $\{f_i\}$
such that $\sum f_i = 1$.

Thus we get

\beqa  \lefteqn{d(r)  \geq \ d_{H_d}(r)  + } \nonumber \\
&&  \sup_{\begin{array}{c}\{(f_1,f_2,\cdots,f_N): \\
           \sum_{i=1}^{N} \ f_i = 1 \} \end{array} }  \ \inf_{\begin{array}{c} \{ (r_1,r_2,\cdots,r_N): \\
                                                               \sum_{i=1}^{N} \ f_i r_i = 2r \} \end{array}}
            \sum_{i=1}^{N} {d_i(r_i)}.  \nonumber \\
            \label{eq:MIMO_NAF_improved}
\eeqa

The scheme in \cite{YanBelMimoAf} is now a special case of this
protocol where all the relays are used for a equal duration of
time, i.e., $f_i = 1/N$ for all $i$. After substituting $\theta_i
:= \frac{r_i}{2Nr}$, we get

\beqn d(r) \geq d_{H_d}(r) +
\inf_{(\theta_1,\theta_2,\cdots,\theta_K): \ \sum_{i=1}^{K} \
\theta_i = 1 }\ \sum_{i=1}^{K} {d_i( 2 N \theta_i r)} \eeqn which
is indeed the formula in \emph{Theorem 2} of \cite{YanBelMimoAf}.
However the lower bound on DMT that we have in
~\eqref{eq:MIMO_NAF_improved} is better than the lower bound in
\emph{Theorem 2} of \cite{YanBelMimoAf} since we allow for
arbitrary periods of activation which is a more general approach.

 \vspace{0.05in} \emph{Example 5: Multiple-Antenna, Multiple-Relay,
Generalized NAF protocol} \vspace{0.05in}

Let us now consider a $N$-relay network with the source and destination having $n_s$ and $n_d$ antennas and the
relays having a single antenna each. For this network, the generalized NAF protocol was proposed in
\cite{RaoHas}, where during the first $T$ time instants, the source transmits to the $N$ relays. Over the next
$T$ time slots, the relays transmit a linear transformation of the vector received over the prior $T$ time
slots. This induces an effective channel matrix between the source and the destination, which will be of the
form:
\beqa \bold{H} & = & \left[ \begin{array}{cc}  \bold{H}_D & 0\\
     \bold{H}_R  & \bold{H}_D \end{array} \right], \label{eq:gen_NAF}\eeqa

where $ \bold{H}_D = I_{T} \otimes \bold{H}_d$,  with $I_T$
denoting the identity matrix of size $T$, $\otimes$ denoting the
tensor product and $\bold{H}_{d}$ denoting the $n_s \times n_d$
fading matrix corresponding to the direct link between the source
and the destination. $\bold{H}_R$ is a $Tn_d \times Tn_s$ matrix
which depends not only on the channel fading coefficients, but
also on the linear transformations employed at the relays
corresponding to the relaying path, which we is the effective
relaying matrix.

Now, $\bold{H}$ is blt and therefore, we invoke
Theorem~\ref{thm:main_theorem} to get, $d_{H}(r) \geq
d_{H^{(0)}}(r) + d_{H^{(\ell)}}(r)$. Now the matrix
$\bold{H}^{(0)}$ corresponds to a block-diagonal matrix with
$\bold{H}_D$ repeated twice along the diagonal or effectively,
$\bold{H}_d$ repeated $2T$ times along the diagonal and clearly
$\bold{H}^{(\ell)} = \bold{H}_R$. Therefore $d_{H^{(0)}}(r) =
d_{H_d}(\frac{r}{2T})$.

The protocol utilizes $2T$ time instants to induce the effective
channel matrix $\bold{H}$ and therefore the DMT of the protocol
$d(r)$ can be given in terms of the DMT of the matrix $\bold{H}$
as $d(r) =
d_{H}(2Tr)$. Thus, \beqa d(r) & = & d_{H}(2Tr) \nonumber \\
& \geq & d_{H^{(0)}}(2Tr) + d_{H^{(\ell)}} (2Tr) \nonumber \\
& = & d_{H^{(0)}}(2Tr) + d_{H_R} (2Tr) \nonumber \\
& = & d_{H_d}(r) + d_{H_R} (2Tr). \label{eq:gen_NAF_mid} \eeqa

We will now present this DMT inequality in the language of
\cite{RaoHas}. In \cite{RaoHas}, the DMT of the effective relaying
matrix, $\bold{H}_R$ is computed after compensating for only a
rate loss of $T$ time instants and let us call this as $d_R(r)$,
i.e., $d_R(r) := d_{H_R} (Tr)$. Let us call the DMT of the direct
link as $d_D(r) := d_{H_d} (r)$. Now \eqref{eq:gen_NAF_mid} can be
re-written as \beqa d(r) \geq d_{D}(r) + d_{R}(2r),
\label{eq:gen_NAF_final} \eeqa which thus proves Conjecture 1 in
\cite{RaoHas}.

\section{Characterization of Extreme Points of DMT of Arbitrary Networks\label{sec:extreme_points}}

In this section, we move on to considering multi-hop networks. We
show that the min-cut is equal to the diversity for arbitrary
multi-terminal networks with multi-antenna nodes irrespective of
whether the relays operate under the half-duplex constraint or
not. We also show for ss-ss full-duplex networks that the maximum
multiplexing gain is equal to the min-cut rank. These two results
put together characterize the two end-points of the DMT of
full-duplex ss-ss networks.

\subsection{Representation of Multi-Antenna Networks \label{sec:representation_multiple}}

In Section~\ref{sec:introduction}, we described how a network is
represented as a graph. The graph-representation of a network
described in Section~\ref{sec:introduction} does not differentiate
between the case with single-antenna nodes and that with
multiple-antenna nodes. We make this distinction in a new
representation of network, described below. We will use this
representation throughout this section.\footnote{A similar
representation for deterministic networks is used in
\cite{AveDigTse1}, albeit in a context different from multiple
antenna nodes. }

Consider a ss-ss wireless network with nodes potentially having
multiple antennas. Every terminal in the network is represented by
a super-node and every antenna attached to the terminal is
represented by a small node associated with the super-node. There
are edges drawn between small nodes of distinct super-nodes,
representing communication channel between antennas of different
terminals. Thus every edge is associated with a scalar fading
coefficient. Since we are dealing with wireless networks, we
assume that the broadcast and interference constraints hold. In
effect the vector $\bold{y}_i$ received by a super-node $i$ with
$m_i$ antennas can be given in terms of the transmitted vectors
$\bold{x}_i$ by \beqn {\bf y_i} = \sum_{j \ \in \
\text{In}(i)}H_{ij} {\bf x_j} + {\bf w_i}, \eeqn where ${\bf y_i}$
and ${\bf w_i}$ are $m_i$ length column column vectors, ${\bf
x_j}$ is a $m_j$ length vector and $H_{ij}$ is a $m_i \times m_j$
transfer matrix between the super-node $i$ and super-node $j$,
containing entries with $\mathbb{C}\mathcal{N}(0, 1)$
distribution. Every cut $\omega$ in the network is associated with
a channel matrix, which we will denote by ${\bf H}_{\omega}$. Fig.
\ref{fig:mincut_multiple_antennas} illustrates this representation
for the case of a single source $S$, two relays $R_1$ and $R_2$
and a sink $D$.

\begin{figure}[h]
  \centering
  \subfigure[Original network with multiple antenna nodes]{\label{fig:mincut_1}\includegraphics[width=50mm]{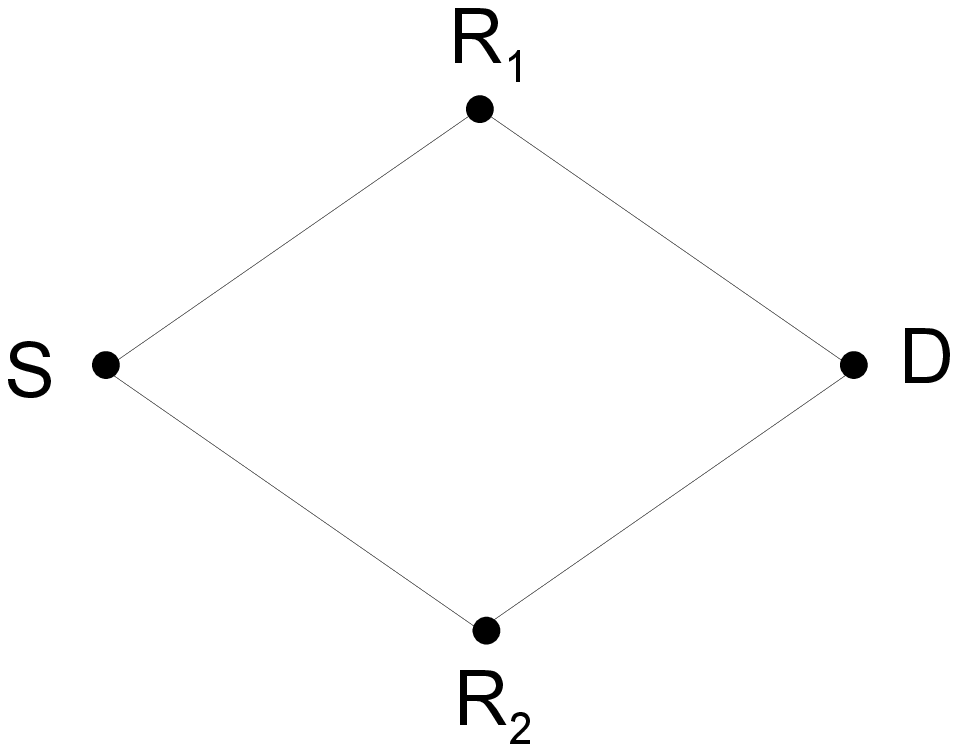}}
  \subfigure[Equivalent network with single antenna nodes]{\label{fig:mincut_2}\includegraphics[width=50mm]{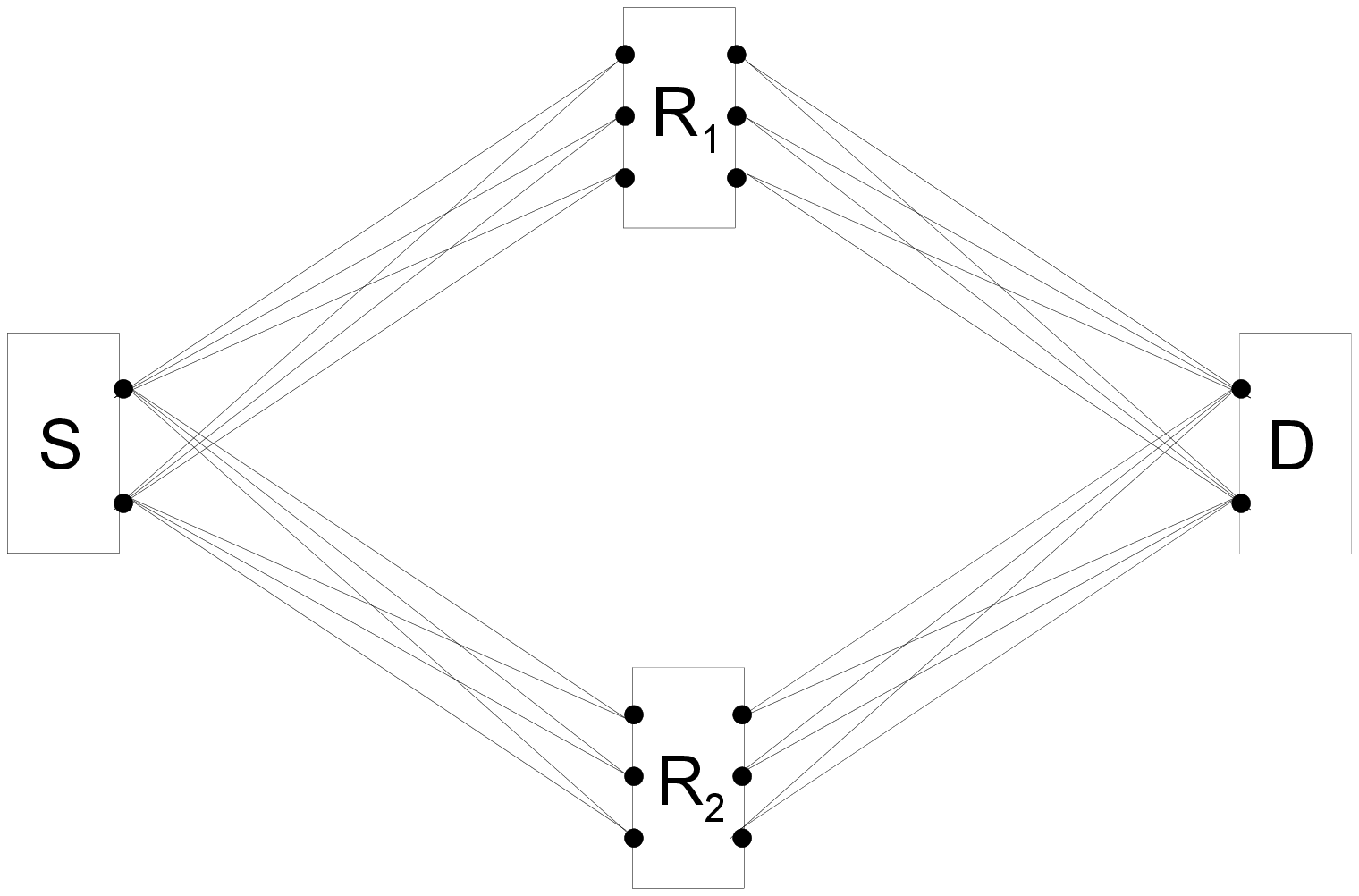}}
  \caption{Source and sink have $2$ antennas each, relays have $3$ each}
  \label{fig:mincut_multiple_antennas}
\end{figure}

It must be noted that even wireline networks can be converted into
the above model of wireless networks. This can be done by adding
as many number of small nodes in a super node as the number of
edges emanating from or arriving at a node. Then, by making the
coefficients of chosen edges to zero (or equivalently by removing
corresponding edges from the representation), the broadcast and
interference constraints can be nullified. Thus the class of
wireline networks are naturally embedded in the class of wireless
networks in the above representation.

\subsection{Min-cut equals Diversity \label{sec:mincut_diversity}}

\bthm \label{thm:mincut} Consider a multi-terminal fading network
with nodes having multiple antennas with edges connecting antennas
on two different nodes having i.i.d. Rayleigh-fading coefficients.
The maximum diversity achievable for any flow is equal to the
min-cut between the source of the flow and the corresponding sink.
Each flow can achieve its maximum diversity simultaneously. \ethm

\bpf We first consider the case where there is only a single
source-sink pair. We will handle the case of single and
multiple-antenna nodes separately. \vspace{0.05in}

\emph{Case I: Network with single antenna nodes}

\vspace{0.05in}

Let the source be $S_i$ and sink be $D_j$. Let $\Lambda_{ij}$
denote the set of all cuts between $S_i$ and $D_j$.

From the cut-set upper bound on DMT (see
Lemma~\ref{lem:CutsetUpperBound}),
\beqan d(r) & \leq & \min_{\Omega \in \Lambda_{ij}}d_{\Omega}(r) \\
\Rightarrow  d(0) & \leq & \min_{\Omega\in
\Lambda_{ij}}d_{\Omega}(0)
\\ & =: & \min_{\Omega\in
\Lambda_{ij}} m_{\omega}, \eeqan where $m_{\omega}$ is the number
of edges crossing from the source side to the sink side in the cut
$\omega$. So, now $d(0) \leq m$, where $m:= \min_{\omega}
m_{\omega}$ is the min-cut.

It suffices to prove that a diversity order equal to $m$ is
achievable. We know from Menger's theorem in graph theory (see for
eg. ~\cite{West}), that the number of edges in the min-cut is
equal to the maximum number of edge-disjoint paths between source
and the sink. Schedule the network in such a way that each edge in
a given edge-disjoint path is activated one by one. The same is
repeated for all the edge-disjoint paths. Let the number of edges
in the $i$th edge-disjoint path be $n_i$. The $j$th edge in the
the $i$th edge-disjoint path is denoted by $e_{ij}$ and the
associated fading coefficient be $\bold{h}_{ij}$. Now define
$\bold{h}_i := \prod_{j=1}^{n_i} \bold{h}_{ij}, i=1,2,...,m$. So
the activation schedule will be as follows: $e_{11}$, $e_{12}$,
$\cdots$, $e_{1(n_1)}$, $e_{21}$, $\cdots$, $e_{2(n_2)}$,
$\cdots$, $e_{m1}$, $e_{m2}$, $\cdots$, $e_{m(n_m)}$, where each
edge is activated one at a time. The total number of time slots
required for the protocol is $N := \Sigma_{i=1}^{m}n_i$. This in
effect creates a parallel channel between the source $S_i$ and
destination $D_j$. The parallel channel contains $m$ links, with
the fading coefficients $\bold{h}_i$ on the link $i$. With this
protocol in place, the equivalent channel seen by a symbol is

\beqan \bold{H} & = & \left[
\begin{array}{cccccc}
        \bold{h}_1         &   0   & \hdots    & &&0\\
        0   & \bold{h}_2   &   0   & &&0\\
        \vdots      &       &\vdots &\ddots &&\\
        0 & \hdots  & \hdots      & && \bold{h}_m
        \end{array}
        \right].
\eeqan

This is a parallel channel with all the channels being independent
of each other and the DMT of the channels being identical.
Therefore we can use Corollary~\ref{cor:parallel_identical} and
obtain the DMT of the parallel channel as \beqa d_H(r) & = &
(m-r)^{+} . \eeqa This DMT can be achieved by using a DMT optimal
parallel channel code.

The protocol utilizes $N$ time instants to induce this effective
channel matrix, and therefore, the DMT of the protocol can be
given in terms of the DMT of the channel matrix as \beqa d(r) & =
& d_H(Nr)
\\ & = & (m-Nr)^{+}. \eeqa

Hence the maximum achievable diversity is $m$.

\vspace{0.1in}

\emph{Case II: Network with multiples antenna nodes}

\vspace{0.1in}

In the multiple antenna case, we pass on to the new representation
described in Section~\ref{sec:representation_multiple}. We regard
any link between a $n_t$ transmit and $n_r$ receive antenna as
being composed of $n_tn_r$ links, with one link between each
transmit and each receive antenna. Note that it is possible to
selectively activate precisely one of the $n_tn_r$
Tx-antenna-Rx-antenna pairs by appropriately transmitting from
just one antenna and listening at just one Rx antenna. As is to be
expected, in this modified representation, a cut is defined as
separating super-nodes into two sets since super-nodes represent
distinct terminals. With this modification, the same strategy as
in the single antenna case can then be applied to achieve a
diversity equal to the min-cut in the network.

\begin{figure}[h]
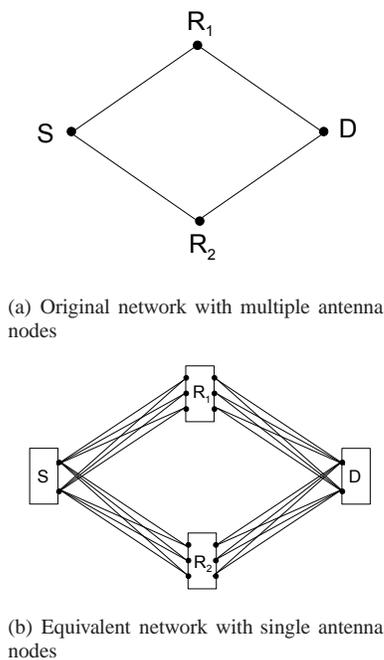

  \centering
  \subfigure[Original network with multiple antenna nodes]{\label{fig:mincut_1}\includegraphics[width=50mm]{mincut_1.eps}}
  \subfigure[Equivalent network with single antenna nodes]{\label{fig:mincut_2}\includegraphics[width=50mm]{mincut_2.eps}}
  \caption{Illustration: $n_S = n_D = 2, n_1 = n_2 = 3$}
  \label{fig:mincut_multiple_antennas}
\end{figure}

Fig.~\ref{fig:mincut_multiple_antennas} illustrates this
conversion for the case of a single source $S$, two relays $R_1$
and $R_2$ and a sink $D$. Having converted the multiple antenna
network into one with single antenna nodes, \emph{Case II} follows
from \emph{Case I}. For the example shown in the figure, the
min-cut and therefore the diversity is equal to $12$.

Thus the proof is complete for the single flow from $S_i$ to
$D_j$.

When there are multiple flows in the network, we simply schedule
the data of all the flows in a time-division manner. This will
entail a rate loss - however, since we are interested only in the
diversity, we can still achieve each flow's maximum diversity
simultaneously. \epf

\subsection{Maximum Multiplexing Gain equals Minimum Rank \label{sec:degrees_of_freedom}}

In this section, we determine the maximum multiplexing gain (MMG)
for multi-antenna ss-ss networks to be equal to the min-cut rank
(which will be formally defined later). For ss-ss networks with
single-antennas, the MMG is lesser than one, because the source
has a single antenna and the cut with source at one side and the
rest of the nodes on the other side will yield an upper bound on
MMG as one. It is possible to attain the optimal MMG of $1$ by
activating one path between the source to the destination either
using amplify-and-forward or a decode-and-forward strategy.
However, the MMG-optimal strategy becomes unclear when the number
of antennas is greater than $1$.

We use results from a recent work on deterministic wireless
networks \cite{AveDigTse2} to arrive at strategies for achieving
the maximum multiplexing gain of a fading network. The
achievability strategies for deterministic wireless networks are
lifted to fading networks using simple algebraic techniques. We
begin with discussing a new representation for ss-ss networks,
potentially having multiple antenna nodes, which will be used in
this section.

\subsubsection{Linear Deterministic Wireless networks \label{sec:lin_det_wnw}}

In defining deterministic\footnote{By deterministic network, we
will always mean linear deterministic network.} wireless networks,
we follow \cite{AveDigTse1}. Every terminal in the network is
represented by a super-node and each node possesses $q$ small
nodes associated with the super-node. All operations take place
over a fixed finite field $\mathbb{F}_p$. There are edges drawn
between small nodes of distinct super-nodes, representing
communication channel between antennas of different terminals.
Since we are dealing with deterministic wireless networks, we
assume that the broadcast and interference constraints hold. In
effect the vector ${\bf y}_i$ received by a super-node $i$ can be
given in terms of the transmitted vectors ${\bf x}_j$ of various
nodes by \beqn {\bf y_i} = \sum_{j \ \in \ \text{In}(i)}G_{ij}
{\bf x_j}, \eeqn where ${\bf y_i}$ and ${\bf x_j}$ are $q$ length
column column vectors in $\mathbb{F}_p$, and $G_{ij}$ is a $m_i
\times m_j$ transfer matrix between the super-node $i$ and
super-node $j$, taking values in $\mathbb{F}_p$. Every cut
$\omega$ in the deterministic network is associated with a channel
matrix, which we will denote by ${G}_{\omega}$.

The network model of linear deterministic networks thus described
has close similarities with representation of ss-ss fading
networks described in Section~\ref{sec:representation_multiple}
with the multiple antennas taking the place of small nodes in the
case of fading networks. The difference between the two are only
that deterministic network has noise-free links in comparison to
the noisy links in the fading case, and that every edge
coefficient is a finite field element in the deterministic
network, in place of complex fading coefficient. In deterministic
networks, each node transmits a $q$-tuple over the finite field.
The theorem below from \cite{AveDigTse1}, computes the capacity
\footnote{We use the term capacity to signify $\epsilon$-error
capacity, as is conventional.} of a ss-ss linear deterministic
wireless network.

\bthm \cite{AveDigTse1} \label{thm:ADT_ss} Given a linear
deterministic ss-ss wireless network over any finite field
$\mathbb{F}_p$, $\forall \ \epsilon
> 0$,  the capacity $C$ of such a relay network
is given by, \beqn C = \min_{\omega \ \in \ \Omega}
\text{rank}(G_{\omega}), \label{eq:ThmLinDetNet} \eeqn where the
capacity is specified in terms of the number of finite field
symbols per unit time. A strategy utilizing only linear
transformations over $\mathbb{F}_p$ at the relays is sufficient to
achieve this capacity. \ethm

\vspace{0.1in} The capacity-achieving strategy in
\cite{AveDigTse1} utilizes matrix transformations of the input
vector received over a period of $T$ time slots at each relay.
This process continues for $L$ blocks, therefore the total number
of time instants required for the scheme is $M:=LT$. The
achievability shows the existence of relay matrices $A_i$ at each
relay node $i \in |\mathcal{V}|$, where $\mathcal{V}$ is the set
of vertices in the graph. $A_i$ is of size $qT \times qT$, and it
represents the transformation between the received vector of size
$qT$ to the vector of size $qT$ that is transmitted.

The multi-cast version of Theorem\ref{thm:ADT_ss} is reproduced
below:

\bthm \cite{AveDigTse1} \label{thm:ADT_multicast} Given a linear
deterministic single-source $D$-sink multi-cast wireless network,
$\forall \ \epsilon > 0$,  the capacity $C$ of such a network is
given by, \beqn \label{eq:ThmLinDetNet} C = \min_{j=1,2,..,D} \
\min_{\omega  \in \ \Omega_j} \text{rank}(G_{\omega}). \eeqn where
$\Omega$ is the set of all cuts between the source and destination
$j$. A strategy utilizing only linear transformations at the
relays is sufficient to achieve this capacity. \ethm

\subsubsection{MMG of ss-ss networks}

The main result of this section is given below. \bthm
\label{thm:DOF_ss} Given a ss-ss multi-antenna wireless network,
with Rayleigh fading coefficients, the MMG of the network is given
by \beqn \label{eq:ThmLinDetNet} D = \min_{\omega \ \in \ \Omega}
\mathbb{R} \mathrm{ank}(\bold{H_{\omega}}) . \eeqn

An amplify-and-forward strategy utilizing only linear
transformations at the relays (that do not depend on the channel
realization) is sufficient to achieve this MMG. \ethm

\bpf \emph{(Outline)} The proof proceeds as follows: \ben \item
First, a converse for the MMG is provided using simple cut-set
bounds. \item Then, we convert the fading network into a
deterministic network with the property that the cut-set bound on
MMG for the fading network is the same as the cut-set bound on the
capacity of the deterministic network. \item We then characterize
the zero-error capacity of the linear deterministic wireless
network. \item Finally, we convert a capacity-achieving scheme for
the deterministic network into a MMG-achieving scheme for the
fading network, which matches the converse. \een \epf

The outline of the proof given above is detailed below. A converse
on the degrees of freedom of a ss-ss fading network is immediate
and is formalized in the following lemma.

\blem \label{lem:ss_converse} Given a ss-ss fading network with
i.i.d. Rayleigh fading coefficients, the MMG, $D$, is upper
bounded by the MMG of every cut:

\beqa D & \leq & \min_{\omega \ \in \ \Lambda} \mathbb{R}
\mathrm{ank}(\bold{H_{\omega}}), \label{eq:Converse} \eeqa where
$\Lambda$ denotes the set of all cuts in the network, and
$\bold{H_{\omega}}$ is the matrix corresponding to the cut $\omega
\in \Lambda$. \elem

Next we proceed to the achievability part of the proof. First, we
convert the wireless fading network into a derived linear
deterministic network.\footnote{It must be noted that the
conversion to deterministic network used here is different from
that used in \cite{AveDigTse2} and \cite{CadJafSha}.} The
construction of the derived deterministic network is described
below. We will show that the zero-error capacity of this derived
deterministic network is lower bounded by the upper-bound on the
MMG of the fading network.

Let the number of edges in the fading network be $N$. Fading
coefficients associated with edges of the network are denoted by
$\bold{h}_1,\bold{h}_2,...,\bold{h}_N$. To construct the derived
deterministic network, consider a deterministic network with the
same topology as that of the original fading network. We take $q$,
the vector length in the deterministic network to be equal to the
maximum number of antennas of any node in the fading network. For
nodes with number of antennas less than $q$, we leave the
remaining nodes unconnected. We still need to decide the finite
field size, $p$, and finite field coefficients on all edges to
completely characterize the equivalent finite-field deterministic
network. We shall denote these finite-field coefficients by
$\xi_i, \ i=1,2,\ldots,N$. We shall consider $\{ \xi_i \}$ as
indeterminates, before values are assigned to them.

For determining the field size $p$ and $\{ \xi_i, \ i=1,2,\ldots,N
\}$, we will impose further conditions. In particular, we will
ensure that the deterministic network will have at least the same
capacity as the upper bound on MMG for the fading network. Due to
the similarity between the expression for capacity in
Theorem~\ref{thm:ADT_ss} and MMG terms in
Lemma~\ref{lem:ss_converse}, above condition can be met by making
sure that, cut-by-cut, the rank of the transfer matrix $(G_\omega)
$ in the deterministic network is at least as large as the
structural rank of the transfer matrix $\bold{H}_\omega$, i.e.,
$\text{rank}(G_\omega) \geq \mathbb{R}
\mathrm{ank}(\bold{H}_\omega)$.

Let us fix a cut $\omega$, and let $r_{\omega} := \mathbb{R}
\mathrm{ank}(\bold{H}_\omega)$ be the structural rank of the
transfer matrix of the cut in the fading network. Then, there
exists a $r_\omega \times r_\omega$ sub-matrix (say
$\bold{H}^{'}_\omega$) of $\bold{H}_\omega$, which has structural
rank $r_\omega$. Consider the same cut on the deterministic
network and find the corresponding $r_\omega \times r_\omega$
sub-matrix $G^{'}_\omega$ of the transfer matrix $G_\omega$. Now
consider the determinant of the matrix $G^{'}_\omega$. The
determinant is a polynomial in several variables $\xi_i,
i=1,2,..,N$ with rational integer coefficients. Let us call this
polynomial as $f_\omega(\xi_1,\xi_2,..,\xi_N)$. This polynomial is
not identically a zero polynomial over $\mathbb{Q}$. This is
because if it had been, then the substitution of $\xi_i = h_i$
will also yields zero irrespective of the choice of $h_i$, making
the determinant zero even for the gaussian case, leading to a
contradiction. Therefore, $f_\omega$ is a non-zero polynomial. We
also observe that the degree of $f_\omega$ in each of the variable
$\xi_i$ is at most one. The lemma below, easily proved using
elementary algebra, shows that it is possible to identify a finite
field $\mathbb{F}_p$ and an allocation to $\{ \xi_i \}$ with
numbers from $\mathbb{F}_p$ such that $f_\omega$ does not vanish.

\blem \label{lem:poly} Given a polynomial
$f(\xi_1,\xi_2,...,\xi_N)$ with integer coefficients, which is not
identically zero, there exists a prime field $\mathbb{F}_p$ with
$p$ large enough, such that the polynomial evaluates to a non-zero
value at least for one assignment of field values to the formal
variables. \elem

However we want ensure the above condition for every cut in the
network. To do so, consider the polynomial \beqa
f(\xi_1,\xi_2,..,\xi_N) := \prod_{\omega  \in \Omega}
f_\omega(\xi_1,\xi_2,..,\xi_N). \eeqa Now, the polynomial $f$ is
non-zero since it is a product of non-zero polynomials $f_\omega$
and the degree of $f$ in any of the variables is at-most
$|\Omega|$. We want a field $\mathbb{F}_p$ and an assignment for
$\xi_i$ from the field such that $f$ is nonzero. Using
Lemma~\ref{lem:poly}, such an assignment exists. Let us choose
that $p$ and the assignment that makes $f$ non-zero. Thus we have
a deterministic wireless network whose capacity is guaranteed to
be greater than or equal to the MMG upper bound, given in
\eqref{eq:Converse}.

Next, we prove that the zero error capacity, $C_{ZE}$, of a linear
deterministic network is equal to its $\epsilon$-error capacity.

\bdefn \cite{Shannon} The zero error capacity of a channel is
defined as the supremum of all achievable rates across the
channels such that the probability of error is exactly zero.
\edefn

\bthm \label{thm:ZEC} The zero error capacity of a ss-ss
deterministic wireless network is equal to \beqan \label{eq:ZEC}
C_{ZE} & = &  \min_{\omega \ \in \ \Omega}
\mathrm{rank}(G_{\omega}) \eeqan This capacity can be achieved
using a linear code and linear transformations in all relays.
\ethm

\bpf We will prove this theorem using the $\epsilon$-error
capacity result from Theorem~\ref{thm:ADT_ss}. Let the ss-ss
deterministic network be composed of $M$ relay nodes. From the
achievability result in the proof of Theorem~\ref{thm:ADT_ss},
given any $\epsilon > 0$ and rate $r < C$, there exists a block
length $T$, number of blocks $L$, set of linear transformations
$A_j, j = 1,2,...,M$ of size $qT \times qT$ used by all relays and
a code book $\mathcal{C}$ for the source, such that the average
probability of error, $P_e$, is less than or equal to $\epsilon$.
Each codeword $x_i \in \mathcal{C}$ is a $qT \times 1$ vector that
specifies the entire transmission from the source. Let
$x_1,...,x_{|\mathcal{C}|}$ be the codewords.

Let us assume that the sink listens for $L^{'} \geq L$ blocks in
general to account for the presence of paths of unequal lengths in
the network between source and sink, (for large $L$, we would have
$\frac{L^{'}}{L} \rightarrow 1$, so this does not affect rate
calculations). Let $M := LT$ and $M^{'} := L^{'}T$. The transfer
equation between the source and the destination vectors are
specified by, $ \bold{y} = G  \bold{x} $ since all transformations
in the network are linear. Here $G$ is a $qM^{'} \times qM$
matrix, $ \bold{x}$ is the $M$-length transmitted vector, and $
\bold{y}$ is the $M^{'}$-length received vector.

Given the transmitted vector $\bold{x}_i$ corresponding to a
message $m_i$ at the source, the decoder either makes an error
always or never makes an error. This is because the channel is a
deterministic linear map, and error is only due to the fact that
$x_i$ and $x_j$ are mapped to the same vector at the decoder. Let
$P_e^{i}$ be the probability of error, conditioned on the fact
that the $i$-th codeword, $ \bold{x}_i$ is transmitted. Then
$P_e^{i} \in \{ 0,1 \}$ according to the argument above and the
average codeword error probability \beqan P_e =
\frac{1}{|\mathcal{C}|} \sum_{i=1}^{ |\mathcal{C}|} P_e^{i} \ \leq
\  \epsilon & \Rightarrow & \sum_{i=1}^{ |\mathcal{C}|} P_e^{i}
\leq \epsilon |\mathcal{C}|. \eeqan This means that at least
$(1-\epsilon)|\mathcal{C}|$ codewords have zero probability of
error. Therefore if we choose only these
$(1-\epsilon)|\mathcal{C}|$ codewords as an expurgated code
$\mathcal{C}^{'}$, then the code has zero probability of error
under the same relay matrices and decoding rule. The rate of the
code is however $\bar{r} = r - \frac{\log \{(1-\epsilon)^{-1}
\}}{M}$. Let $\delta = \frac{\log \{(1-\epsilon)^{-1}\}}{M}$ be
the rate loss and therefore, the expurgated code has negligible
rate loss as $M$ becomes large. Now, we have established a zero
error code of rate $r - \delta$. By choosing $r$ arbitrarily close
to $C$ and $M$ large, we get $C_{ZE} = C$.

The code ${\cal C}$, as given in \cite{AveDigTse1} is non-linear,
and so is the case of ${\cal C}^{'}$. However, we can obtain a
linear code with zero probability of error. Since there exists a
zero error code for rate $\bar{r}$ with block length $T$ and
number of blocks $L$, it means that the transfer matrix $G$
between the source and the sink has rank at least $\bar{r}M$.
Hence $G$ has a sub-matrix $G^{'}$ of size $\bar{r}M \times
\bar{r}M$, which is of full rank. By activating appropriate nodes,
we can obtain the effective transfer matrix to be $G^{'}$. In that
case, a linear code of rate $\bar{r}$ which communicates only on
the $ \bar{r} M$ subspaces can be used to achieve zero error. \epf

Thus, for a given fading network, we have constructed an
equivalent deterministic network. In the equivalent network, we
also have a zero-error achievable rate $\bar{r}$ using a linear
code of block length $T$ and $L$ blocks with linear strategies at
the relays. This achievable rate is related to the MMG of the
original fading network as follows:
\beqan \bar{r} + \delta & = & C_{ZE} \\
& = & \min_{\omega \ \in \ \Omega} \mathrm{rank}(G_{\omega}) \\
& \geq & \min_{\omega \ \in \ \Omega} \mathbb{R}
\mathrm{ank}(\bold{H_{\omega}}) \eeqan Further, the positive
constant $\delta$ can be made as small as possible as we wish by
increasing the block length $T$. Now, when we use the zero-error
scheme detailed above, the transfer matrix $G$ of size $qM \times
qM$ between the input and the output vectors $\bold{x}$ and
$\bold{y}$ is at least of rank $\bar{r}M \approx CM$.

Finally, we lift the achievability strategy of zero-error capacity
in the equivalent deterministic networks to arrive at an
achievable strategy for MMG in corresponding fading network.

In the reduced deterministic network of a fading network, to
achieve the zero-error capacity, the relays perform matrix
operations $A_i$ on received vectors for $T$ time durations. Since
each received vector is of size $q$, the matrix $A_i$ is of size
$qT \times qT$. Now we use the same strategy for the fading
network, i.e., all relays use the same matrices $A_i$, that are
obtained via the zero-error strategy in the reduced deterministic
network. Though the entries of $A_i$ belong to $\mathbb{F}_p$,
they can be treated as integers by identifying the elements of
$\mathbb{F}_p$ with the integers $0,1,,...,(p-1)$. Therefore the
matrices $A_i$ can also be interpreted as matrices over
$\mathbb{C}$. By using linear maps $A_i$ at relays in the fading
network, we get an induced channel matrix ${\bf H}$, and effective
channel would be of the form, ${\bf y} \ = \ {\bf H}{\bf x}  +
{\bf w}$. As is shown in Theorem~\ref{thm:DOF_rank}, MMG offered
by this channel is equal to $\mathbb{R} \mathrm{ank} (\bold{H})$.
We shall prove that MMG offered by this induced channel is greater
than or equal to $\bar{r}M$, i.e., to show that $\mathbb{R}
\mathrm{ank}(\bold{H}) \geq \bar{r}M$. That is equivalent to show
that there exists an assignment of $\bold{h}_i = h_i$ in the
fading network such that $\text{rank}(H) \geq \bar{r}M$.

In the proof of Theorem~\ref{thm:ZEC}, we restricted the operation
of the derived deterministic network to create a transfer matrix
$G$ of size $qM \times qM$ with rank greater than or equal to
$\bar{r}M$. Now we have a similar transfer matrix $\bold{H}$ in
the fading network. If we assign the underlying random variables
$\bold{h}_i$ to be equal to $\xi_i$, again by identifying the
elements of $\mathbb{F}_p$ with the integers $0,1,,...,(p-1)$, we
have an assignment of $\bold{H}$ that has rank at least
$\bar{r}M$. Since the strucutral rank is the maximum possible rank
under any assignment, we get that, \beqan \mathbb{R} \text{ank}
(\bold{H}) & \geq & \text{rank}(G) \geq \bar{r}M. \eeqan

The induced channel therefore has a MMG equal to $\bar{r}M$ by
Theorem~\ref{thm:DOF_rank}. Since the network is operated for $M =
LT$ time slots in order to obtain a MMG greater than or equal to
$\bar{r}M$, the MMG of the network per time slot is greater than
or equal to $\bar{r}$. By increasing the block length $T$ and the
number of blocks $L$, the achievable MMG can be made arbitrarily
close to $C_{ZE}$. Thus the upper bound given in
Lemma~\ref{lem:ss_converse} is achieved, and hence MMG of ss-ss
fading network is given by \beqan D & = & \min_{\{ \omega \ \in \
\Lambda\}} \mathbb{R} \mathrm{ank}(\bold{H}_{\Lambda}). \eeqan

\subsection{MMG for Multi-casting \label{sec:dof_multicast}}

In this section, we extend the result on MMG to the multi-casting
scenario.

\bthm \label{thm:DOF_multicast} Given a single-source $D$-sink
multi-cast gaussian wireless network,
 with Rayleigh fading coefficients, the MMG of the network is given by
\begin{eqnarray}
\label{eq:ThmLinDetNet} D = \min_{ \{j=1,2,...,D \}} \
\min_{\omega \ \in \ \Omega_j} \mathbb{R}
\mathrm{ank}(\bold{H}_{\omega}).
\end{eqnarray}
An amplify-and-forward strategy utilizing only linear
transformations at the relays is sufficient to achieve this MMG.
\ethm

\bpf The proof uses Theorem~\ref{thm:ADT_multicast}, goes in the
similar lines of that of Theorem~\ref{thm:DOF_ss} and is omitted
here for brevity. \epf

\section{DMT Bounds for Single Antenna Relay Networks \label{sec:full_duplex}}

In this section, we consider ss-ss networks equipped with
full-duplex single-antenna nodes. We provide a lower bound to the
DMT of such a network by exploiting Menger's theorem.

\bdefn Consider a network $N$ and a path $P$ from source to sink. This path $P$ is said to have a {\em shortcut}
if there is a single edge in $N$ connecting two non-consecutive nodes in $P$. \edefn

\bthm \label{thm:FD_No_Direct_Path} Consider a ss-ss full-duplex
network with single antenna nodes. Let the min-cut of the network
be $M$. Let the network satisfy \emph{either} of the two
conditions below: \ben \item The network has no directed cycles,
or \item There exist a set of $M$ edge-disjoint paths between
source and sink such that \emph{none} of the $M$ paths  have
shortcuts. \een Then, a linear DMT $d(r) = M(1-r)^{+}$ between a
maximum multiplexing gain of $1$ and maximum diversity $M$ is
achievable. \ethm

\bpf Given that the network has min-cut $M$, there are $M$
edge-disjoint paths from source to sink by Menger's theorem
\cite{West}. Let us label the edge-disjoint paths
$e_1,e_2,...,e_M$. Let the product of the fading coefficients
along the path $e_i$ be $g_i$. Let $D_i$ be the delay of each
path. Let $D = \max D_i$. We add delay $D-D_i$ to the path $e_i$
such that all paths now are of equal delay. We activate the edges
as follows:

\ben \item Activate all edges in the edge-disjoint path $e_1$
simultaneously for a period $T$, where $T >> D$. The source, on
the first $T-D$ activations, will transmit $(T-D)$ coded
information symbols, followed by a sequence of $D$ zero symbols.
The reason for this will become clear shortly. The net effect will
be to create a $((T-D) \times (T-D)$ transfer matrix ${\bf H}_1$
from the $(T-D)$ source symbols to the last $(T-D)$ symbols
received by the destination (the first $D$ symbols received by the
destination are all zero).

The matrix ${\bf H}_1$ will be either upper-triangular or
lower-triangular, with the elements along the diagonal all equal
to the path gain ${\bf g}_1$ on path $e_1$, according to whether
the condition $1$ or condition $2$ of the theorem is satified.
First, we explain the case when the graph has no directed cycles.
In this case, off-diagonal terms above the diagonal can arise due
to the presence of short-cuts. However, off-diagonal terms below
the diagonal would constitute a directed cycle and will thus not
appear. Therefore the matrix will be upper-triangular in this
case. Next, for the case when the graph has no shortcuts,
off-diagonal terms below the diagonal can arise due to the
presence of cycles in the graph. However, no terms above the
diagonal will be present because of the presumed absence of
shortcuts. Thus the induced matrix will be lower-triangular in
structure.

\item Repeat \emph{Step 1} for all edge-disjoint paths
$e_2,...,e_M$. The net transfer matrix ${\bf H}$ will be block
diagonal of the form \beqa {\bf H} & = & \left[
\begin{array}{cccc}
 {\bf H}_1 & & & \\
  & {\bf H}_2 & & \\
   &  & \ddots &  \\
  & & & {\bf H}_M \end{array} \right] .
  \eeqa
composed of $M$ blocks along the diagonal, one corresponding to each path, and either all of them are upper
triangular or all are lower triangular by the argument above. \een

Now, if $d(r)$ is the DMT of the network operating under the
protocol given above, then \beqa d(r) & = & d_H(MTr), \eeqa since
$MT$ time instants were used up by the protocol in order to obtain
the indcued channel matrix $\bold{H}$. We next proceed to lower
bound $d_H(r)$. By Theorem~\ref{thm:main_theorem}, we have the
lower bound \beqa d_H(r) & \geq & d_{H^{(0)}}(r) \eeqa where ${\bf
H}^{(0)}$ is the matrix corresponding to the diagonal terms in
${\bf H}$. Next, we observe that ${\bf H}^{(0)}$ corresponds to a
parallel channel with $M$ fading coefficients ${\bf g}_1,{\bf
g}_2,\cdots,{\bf g}_M$, each of them repeated $T-D$ times. We can
compute the DMT of this parallel channel using
Lemma~\ref{lem:parallel_dependent}. Thus we get, \beqa
d_{H^{(0)}}(r) & = & M \left( 1-\frac{r}{M(T-D)} \right)^{+} \\
\Rightarrow d(r) & = & d_{H} (MTr) \geq  d_{H^{(0)}} (MTr) \eeqa
\beqa d(r) & = & M \left( 1-\frac{MTr}{M(T-D)} \right)^{+}. \eeqa

For $T$ tending to $\infty$, we get $d(r) \geq M(1-r)^{+}$. \epf

\appendices

\section{Proof of Lemma~\ref{lem:tail_probability}\label{app:tail_probability}}

Let the multinomial $f$ be written as a sum of $S$ monomials
$f(X_1,X_2,...,X_M) := \sum_{i = 1}^{S} c_if_i(X_1,X_2,...,X_M$),
where for every $i$, $c_i$ is a constant and $f_i$ is a monomial,
i.e., is comprised only of product of powers of $X_i$. Then for
every assignment, $X_i = h_i$, \beqn |f(h_1,h_2,...,h_M)| \leq
\sum_{i} |c_i||f_i(h_1,h_2,...,h_M)|. \eeqn
Now we have, \beqa \lefteqn{\Pr \{ |f(\bold{h}_1,\bold{h}_2,...,\bold{h}_M)|^2 > k\} } \nonumber \\
& = & \Pr\{|f(\bold{h}_1,\bold{h}_2,...,\bold{h}_M)| > k^{\frac{1}{2}}\} \nonumber \\
& \leq & \Pr \{ \sum_i
|c_i||f_i(\bold{h}_1,\bold{h}_2,...,\bold{h}_M)| > k^{\frac{1}{2}}
\} \label{eq:noise_white_l1_start}
\\
& \leq &  \Pr\left\{\bigcup_{i} \left(
|c_i||f_i(\bold{h}_1,\bold{h}_2,...,\bold{h}_M)|
> \frac{k^{\frac{1}{2}}}{S} \right)
\right\} \nonumber \\
& \leq &  \sum_{i} \Pr\left\{  |c_i|^2|f_i(\bold{h}_1,\bold{h}_2,...,\bold{h}_M)|^2 > \frac{k}{S^2} \right\} \nonumber \\
& \leq &  \sum_{i} \Pr\left\{
|f_i(\bold{h}_1,\bold{h}_2,...,\bold{h}_M)|^2 >
\frac{k}{c_{max}S^2} \right\}, \label{eq:noise_white_l1_mid}\eeqa

where $c_{max}$ is the maximum over all $\{|c_i|^2\}$. Now
$|f_i(h_1,h_2,...,h_M)|^2$ is a monomial in $|h_i|^2$ as well.
Define, $\bold{u}_j := |\bold{h}_j|^2$. Then $\bold{u}_j$ is the
squared norm of a $\mathbb{C}\mathcal{N}(0,1)$ random variable
$\bold{h}_j$, and therefore has an exponential distribution.  We
will regard $|f_i(\bold{h}_1,\bold{h}_2,...,\bold{h}_M)|^2$ as a
monomial $g_i(\bold{u}_1,\bold{u}_2,...,\bold{u}_M)$ in
$\{\bold{u}_j\}$. Thus

\beqan g_i(\bold{u}_1,\bold{u}_2, \ldots, \bold{u}_M) =
\prod_{j=1}^{M} {\bold{u}_j}^{a_{ij}}, \eeqan where $0 \leq a_{ij}
\leq D$ is an integer, where $D$ is the maximum degree of any of
the monomials $g_i$ in any of the variables $\bold{h}_i$.

Now gathering a single term in the summation in RHS of
\eqref{eq:noise_white_l1_mid}, we have, \beqa \lefteqn{\Pr\left\{
|f_i(\bold{h}_1,\bold{h}_2,...,\bold{h}_M)|^2 >
\frac{k}{c_{max}S^2} \right\} } \nonumber \\
& = & \Pr\left\{ g_i(\bold{u}_1,\bold{u}_2, \ldots, \bold{u}_M) > \frac{k}{c_{max}S^2} \right\}  \nonumber \\
& \leq & \Pr\left\{ \bigcup_{j} \left( {\bold{u}_j}^{a_{ij}} >
\left(\frac{k}{c_{max}S^2}\right)^\frac{1}{M} \right) \right\}
\nonumber \eeqa \beqa & \leq & \sum_{j} \Pr\left\{
{\bold{u}_j}^{a_{ij}}
>
\left(\frac{k}{c_{max}S^2}\right)^\frac{1}{M} \right\} \nonumber \\
& = & \sum_{j} \Pr\left\{ {\bold{u}_j} >
\left(\frac{k}{c_{max}S^2}\right)^\frac{1}{Ma_{ij}} \right\}
\nonumber \eeqa \beqa & \leq & \sum_{j} \Pr\left\{ {\bold{u}_j} >
\left(\frac{k}{c_{max}S^2}\right)^\frac{1}{MD}
\right\} \label{eq:tail_prob_power} \\
& = & M \ \Pr\left\{ {\bold{u}_j} >
\left(\frac{k}{c_{max}S^2}\right)^\frac{1}{MD}
\right\} \nonumber \\
& = & M \
\exp\left(-\left(\frac{k}{c_{max}S^2}\right)^\frac{1}{MD}\right)
\label{eq:noise_white_l1_end}. \eeqa Equation
\eqref{eq:tail_prob_power} follows if \beqan  {k}  & \geq &
{c_{max}S^2}. \eeqan We get this condition by setting $\delta :=
{c_{max}S^2} > 0$, since by the hypothesis of the lemma, we have
$k \geq \delta$.

Combining \eqref{eq:noise_white_l1_mid} and
\eqref{eq:noise_white_l1_end}, we get, for $ k \geq \delta$,
\beqan \lefteqn{\Pr \{ |f(\bold{h}_1,\bold{h}_2,...,\bold{h}_M)|^2 > k \}} \\
& \leq & M S \ \exp\left(-\left(\frac{k}{c_{max}S^2}\right)^\frac{1}{MD}\right) \\
& = & A \ \exp\left(-Bk^{\frac{1}{d}}\right) \ \eeqan where $A :=
MS$, $B := \left({\frac{1}{c_{max}S^2}}\right)^{\frac{1}{MD}}$,
$\delta = {c_{max}S^2}$ and $d:=MD$.

\section{Proof of Theorem~\ref{thm:noise_white}\label{app:noise_white}}

\beqan \lefteqn{Pr(\log\det(I+ \rho
\bold{H}\bold{H}^{\dagger}\Sigma^{-1}) \leq r\log\rho) } \\
& \doteq & Pr(\log\det(I+ \rho \bold{H}\bold{H}^{\dagger}) \leq
r\log\rho). \eeqan

Let the correlation matrix of the noise vector be denoted by
$\bold{\Sigma}$. The noise covariance matrix $\bold{\Sigma}$
depends on the channel realization and is therefore a random
matrix, given by,
\beqa \bold{\Sigma} & = & \mathbb{E}_{\bf z}[{\bf z}{{\bf z}^\dagger}] \nonumber \\
& = & I + \sum_{i=1}^{M} \bold{G}_i\bold{G}_i^\dagger .
\label{eq:sigma}\eeqa Let $\lambda_i(A)$, $\lambda_{max}(A)$ and
$\lambda_{min}(A)$ denote the $i^{\text{th}}$ largest, maximum and
minimum eigenvalues of a positive semi-definite matrix $A$. If the
context is clear, we may avoid specifying the matrix, and just use
$\lambda_i$, $\lambda_{max}$ and $\lambda_{min}$ respectively.

To prove the lemma, we will use the Amir-Moez bound on the eigen
values of the product of Hermitian, positive-definite matrices
\cite{AmiMoe}. By this bound, for any two positive definite $n
\times n$ Hermitian matrices $A, B$: \beqa
\lambda_i(A)\lambda_{min}(B) \leq & \lambda_i(AB) & \leq
\lambda_i(A)\lambda_{max}(B). \nonumber \eeqa So we get,  \beqa
\det (I + \rho A B) & = & \prod_{i}(1+\rho\lambda_i({AB}))
\nonumber \\
 & \leq & \prod_{i}(1+\rho\lambda_i(A)\lambda_{{\max}}(B)) \nonumber \\
 & = & \det (I + \rho  \lambda_{{\max}}(B)A ). \nonumber \eeqa
Similarly, \beqan \det (I + \rho A B) & \geq & \det (I + \rho
\lambda_{\min}(B) A). \eeqan Therefore, for any two positive
definite $n \times n$ Hermitian matrices $A, B$,\beqa
\lefteqn{\det (I + \rho \lambda_{\min}(B) A)} \nonumber \\  \ & &
\leq \det (I + \rho A B) \leq  \ \det (I + \rho \lambda_{\max}(B)
A). \label{eq:reduced_Amir_Moez} \eeqa

Applying \eqref{eq:reduced_Amir_Moez} to $A
=\bold{H}\bold{H}^\dagger$ and $B = \bold{\Sigma}^{-1}$, we get \beqa \Rightarrow \lefteqn{\det(I+\rho{\bf H}{\bf H}^\dagger \lambda_{min} ( \bold{\Sigma}^{-1}))} \nonumber \\
& & \leq  \det(I+\rho{\bf H}{\bf H}^\dagger  \bold{\Sigma}^{-1}) \label{eq:amir_moez_lower_bound} \\
 & & \leq \det(I+\rho{\bf H}{\bf H}^\dagger \lambda_{max}
( \bold{\Sigma}^{-1})). \label{eq:amir_moez_upper_bound} \eeqa
Continuing from \eqref{eq:amir_moez_lower_bound} and
\eqref{eq:amir_moez_upper_bound}, we have
\beqa \lefteqn{\Pr\{\log(\det(I+\rho{H}{H}^\dagger \lambda_{min} ( \bold{\Sigma}^{-1}))) < r\log\rho\} } \nonumber \\
& & \geq  \Pr\{\log\det(I+\rho{H}{H}^\dagger  \bold{\Sigma}^{-1})
< r\log\rho\} \nonumber \\
& & \geq \Pr\{\log(\det(I+\rho{H}{H}^\dagger \lambda_{max} (
\bold{\Sigma}^{-1}))) < r\log\rho\}. \nonumber
\\\label{eq:outage_bound} \eeqa
In the following, we will prove that both the bounds coincide as
$\rho$ $\rightarrow$ $\infty$. We begin with the bounds on
$\lambda_{min}( \bold{\Sigma})$ and $\lambda_{max}(
\bold{\Sigma})$. In order to show that the lower and the upper
bounds on the expression converge to the value
$\Pr\{\log(\det(I+\rho{H}{H}^\dagger)) < r\log\rho\}$, we need to
provide a lower bound for each $\lambda_i( \bold{\Sigma})$. Let
$e_i$ be the eigen vector corresponding to $\lambda_i({\Sigma})$
for every realization $\Sigma$ of $ \bold{\Sigma}$. Then, \beqa
{\lambda_i(\Sigma)}\parallel{e_i}\parallel^{2}
 & = & {e_i}^\dagger\Sigma{e_i} \nonumber \\
 & = & {e_i}^\dagger\left(I + \sum_{i=1}^{M} \bold{G}_i\bold{G}_i^\dagger\right){e_i} \nonumber \\
 & = & \parallel{e_i}\parallel^2 + {e_i}^\dagger{\left(\sum_{i=1}^{M} \bold{G}_i\bold{G}_i^\dagger\right)}{e_i} \nonumber \\
 & \geq & \parallel{e_i}\parallel^2 \nonumber \\
\Rightarrow {\lambda_i(\bf \Sigma)} & \geq & 1 \ \ \forall i \nonumber \\
\Rightarrow \lambda_{min}(\bf \Sigma) & \geq & 1 \label{eq:eigen_lower} \\
\Rightarrow \lambda_{max}(\bf \Sigma^{-1}) & \leq & 1 \nonumber
\eeqa \beqa \text{Hence, }
\lefteqn{\Pr\{\log\det(I+\rho \bold{H}\bold{H}^\dagger {\bf \Sigma}^{-1}) < r\log\rho\}} \nonumber \\
& \geq & \Pr\{\log\det(I+\rho \bold{H} \bold{H}^\dagger\lambda_{max}({\bf \Sigma}^{-1})) < r\log\rho\} \nonumber \\
& \geq &  \Pr\{\log\det(I+\rho \bold{H} \bold{H}^\dagger) < r\log\rho\} \\
& \doteq & \rho^{d_{out}(r)}. \label{eq:outage_lower_bound}\eeqa

Now we proceed to get an upper bound on $\lambda_{max}(
\bold{\Sigma})$: \beqa \lambda_{max}({\bf \Sigma}) & = & \lambda_{max}(I + \sum_{i=1}^{M} {\bf G}_i{\bf G}_i^\dagger) \nonumber \\
& = & 1 + \lambda_{max}(\sum_{i=1}^{M}
\bold{G}_i\bold{G}_i^\dagger) \nonumber \eeqa i.e.,
\beqa  \lambda_{max}({\bf \Sigma}) & \leq & 1 + \text{Tr} \ (\sum_{i=1}^{M}\bold{G}_i\bold{G}_i^\dagger) \nonumber \\
& = & 1 + \sum_{i=1}^{M}\text{Tr} \ (\bold{G}_i\bold{G}_i^\dagger) \nonumber \\
& = & 1 + \sum_{i=1}^{M}{||\bold{G}_i||^2}_{F} \nonumber \\
& = & 1 + \sum_{i=1}^{M} \sum_{j=1}^{N^2}|f_{ij}|^2,
\label{eq:dof_poly_bound}\eeqa where $f_{ij}$ represents a
polynomial entry of the matrix $G_i$. Let
$\bold{x}_1,\bold{x}_2,...,\bold{x}_{2L}$ denote in some order,
the real and imaginary parts of
$\bold{h}_1,\bold{h}_2,...,\bold{h}_{L}$. Then, the right hand
side in \eqref{eq:dof_poly_bound} is a polynomial in the
variables, $\bold{x}_i, i=1,2,...,2L$.

This leads to the following inequality, \beq \lambda_{max}(
\bold{\Sigma}) \leq g(\bold{x}_1,\bold{x}_2, \ldots,
\bold{x}_{2L}) + 1, \label{eq:max_lambda_bound_1}\eeq where
$g(x_1,x_2, \ldots, x_{2L})$ is a polynomial without constant term
in the variables $\{ x_i \}$. Let us invoke
Lemma~\ref{lem:tail_probability} for the polynomial $g$ which does
not possess any constant term. The lemma is valid for all $k \geq
\delta$, where $\delta$ depends on $g$. Let us choose $\rho_0$
such that $\rho_0^{\epsilon}-1 \geq \delta$ and therefore $\forall
\rho
> \rho_0$, we have that $\rho^{\epsilon}-1 \geq \delta$. Now, $\forall
\rho > \rho_0$, we can invoke Lemma~\ref{lem:tail_probability} and
get, \beqa \Pr\{ \lambda_{max}( \bold{\Sigma}) > \rho^\epsilon \}
& \leq & \Pr\{ g(\bold{x}_1,\bold{x}_2, \ldots, \bold{x}_S) >
{\rho}^{\epsilon} - 1 \} \nonumber \\
& \leq & A\exp(-B{(\rho^\epsilon - 1)}^{\frac{1}{d}}),
\label{eq:nw_prob_ub} \eeqa for some constants $A$, $B$, $d >0$.

Let $\mathcal{H}$ denote the set of all the fading coefficients in
the network, and let $\hbar \in \mathcal{H}$ denote a realization
of the fading coefficients. Thus $\hbar$ will be a vector
specifying $h_i, i=1,2,...,L$. Clearly, once a $\hbar$ is given,
the values of the matrices $H$ and $G_i$ are all well defined,
since all of them depend only on $h_i$.

Let ${A} = \{ \hbar \in \mathcal{H} \mid \log\det( I +
{\rho}HH^{\dagger}\Sigma^{-1} ) < {\rho}^r \}$ and $B = \{ \hbar
\in \mathcal{H} \mid \lambda_{max}(\Sigma) > {\rho}^{\epsilon} \}$
be two events. Then, \beqa
\Pr(A) & = & \Pr(A \cap B^c) + \Pr(A \cap B) \nonumber \\
& \leq & \Pr(A \cap B^c) + \Pr(B) \nonumber \\
& \dot \leq & \Pr(A \cap B^c) + A\exp(-B{(\rho^\epsilon -
1)}^{\frac{1}{d}}) \label{eq:nw_ab_ub}
\\ \text{Now, } A & = & \{ \hbar \in \mathcal{H} \mid \log\det( I +
{\rho}HH^{\dagger}\Sigma^{-1} ) < {\rho}^r \} \nonumber \\
& \subset & \{ \hbar \in \mathcal{H} \mid \log\det(
I + {\rho}HH^{\dagger}\lambda_{min}(\Sigma^{-1}) ) < {\rho}^r \} \nonumber \\
& = & \{ \hbar \in \mathcal{H} \mid \log\det( I +
{\rho}HH^{\dagger}{(\lambda_{max}(\Sigma))}^{-1} ) < {\rho}^r \} \nonumber \\
A \cap B^c & \subset & \{ \hbar \in \mathcal{H} \mid \log\det( I +
{\rho}^{1-\epsilon}HH^{\dagger} ) < {\rho}^r \}.
\label{eq:nw_abc_ub} \eeqa

Substituting \eqref{eq:nw_abc_ub} and \eqref{eq:nw_prob_ub} in
\eqref{eq:nw_ab_ub}, taking logarithms and dividing by $\log \rho$
on both sides, we have,
 \beqa
\lefteqn{\frac{\log \Pr(A)}{\log\rho}}  \nonumber \\
& \leq & \frac{\log[ \Pr(A \cap B^c) + \Pr(B) ]}{\log\rho} \nonumber   \\
& \dot \leq & \frac{1}{\log\rho} \ . \  \log \ [ \ \Pr\{\hbar \in \mathcal{H} \mid \log\det( I + {\rho}^{1-\epsilon}HH^{\dagger} ) < {\rho}^r\}  \nonumber \\
& \  \  \  & \ \ \ \ \ \  \ \ \ \ \ \ \ \ \ \ + \ \
A\exp(-B{(\rho^\epsilon - 1)}^{\frac{1}{d}}) \ ].
\label{eq:nw_doteq} \eeqa \beqan
\lefteqn{\lim_{\rho\rightarrow\infty} \frac{\log \Pr(A)}{\log\rho}} \\
& \dot \leq & \lim_{\rho\rightarrow\infty} \frac{\log[ \Pr\{\hbar
\in \mathcal{H} \mid \log\det( I + {\rho}^{1-\epsilon}HH^{\dagger}
) < {\rho}^r\} ]}{\log\rho}. \eeqan The last equation follows from
\eqref{eq:nw_doteq}, since the first term in the RHS of
\eqref{eq:nw_doteq} varies inversely with an exponent of $\rho$
whereas the second term is exponential in $\rho$, and therefore
the sum is dominated by the first term. After making the variable
change, $\rho^{'} = \rho^{1-\epsilon}$ and replacing $\rho$ by
$\rho^{'}$, we get \beqa \lefteqn{\Rightarrow
\lim_{\rho\rightarrow\infty} \frac{\log \Pr\{ \log\det( I
+ {\rho} \bold{H}\bold{H}^{\dagger} \bold{\Sigma}^{-1} ) < {\rho}^r \}}{\log\rho}} \nonumber \\
& \dot \leq & (1-\epsilon) \nonumber \\
& & \lim_{\rho\rightarrow\infty} \frac{\log  \Pr\{ \log\det(  I +
{\rho} \bold{H}\bold{H}^{\dagger} ) <
{\rho}^{(\frac{r}{1-\epsilon})}
\}}{\log\rho} \nonumber \\
& \doteq &
(1-\epsilon)\rho^{d_{out}\left(\frac{r}{1-\epsilon}\right)} .
\label{eq:outage_upper_bound} \eeqa In
\eqref{eq:outage_upper_bound}, $\epsilon$ is arbitrary, and we let
it tend to zero. Hence, by \eqref{eq:outage_upper_bound} and
\eqref{eq:outage_lower_bound}, the exponents for both the bounds
in \eqref{eq:outage_bound} coincide and we obtain, \beqan
\lefteqn{\Pr\{\log\det(I+\rho \bold{H}
\bold{H}^\dagger\Sigma^{-1}) < r\log\rho\}} \\ & \doteq &
\Pr\{\log\det(I+\rho \bold{H} \bold{H}^\dagger) < r\log\rho\}.
\nonumber \eeqan

This proves the assertion of the theorem.

\section{Proof of Lemma~\ref{lem:poly_finite_regions}
\label{app:poly_finite_regions}}

Consider any two intervals $R_1, R_2 \in {\cal R}$. If we are not
able to find two such intervals, then clearly $L \leq 1$, and we
are done. Let $R_1 = [a, b]$ and $R_2=[c,d]$, and without loss of
generality assume that $a < b < c < d$, since they are, by
hypothesis, disjoint. First, we claim that there exists a point $b
\leq x_0 \leq c$, such that either $p^{\prime}(x_0) = 0$, or
$p^{\prime\prime}(x_0) = 0$. We now proceed to prove this claim.

Clearly, either of the two conditions
\eqref{eq:poly_local_condition_1} or
\eqref{eq:poly_local_condition_2} is violated just to the right of
the point $x=b$, else, the interval would extend beyond $b$.  We
consider two cases.

\emph{Case 1:} $|p(b)| \geq k$.

\begin{figure}[h]
  \centering
  \subfigure[Case 1(a)]{\label{fig:polynomial_1}\includegraphics[width=75mm]{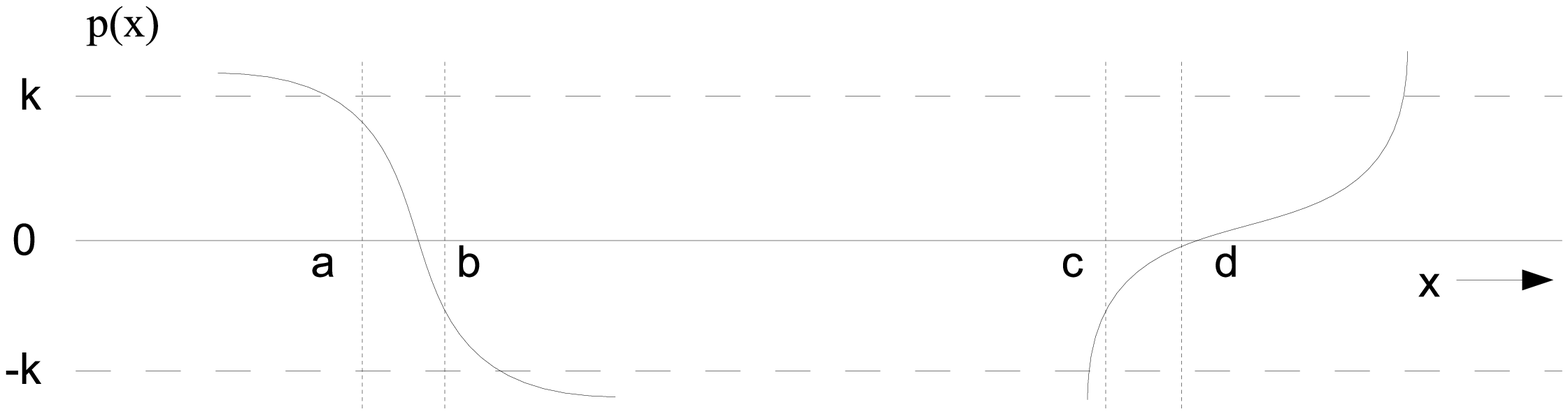}}
  \subfigure[Case 1(b)]{\label{fig:polynomial_2}\includegraphics[width=75mm]{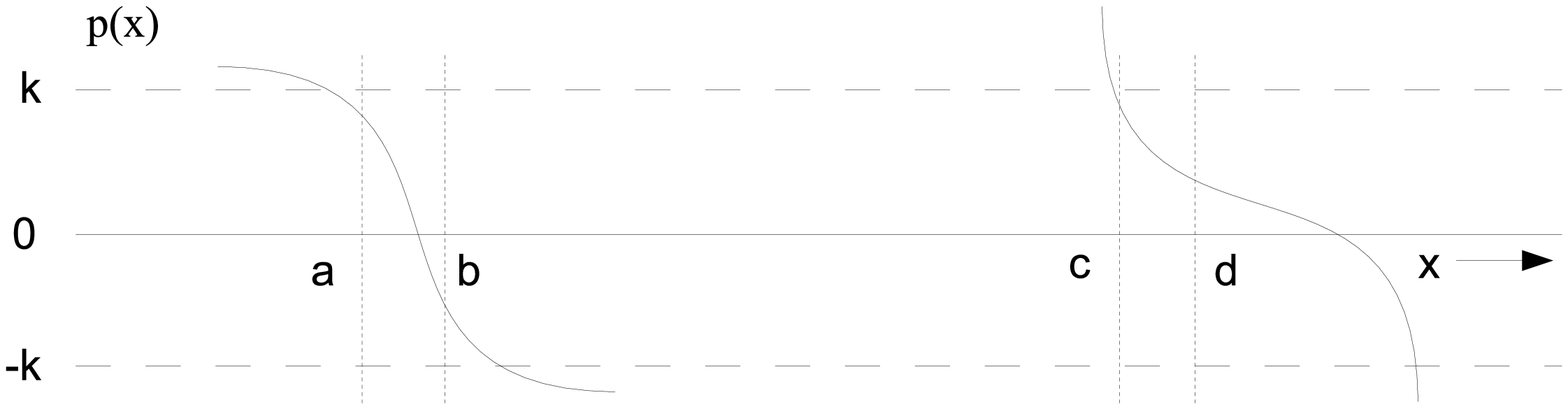}}
  \caption{Condition \eqref{eq:poly_local_condition_1} is violated in $[b, c]$ }
  \label{fig:polynomial_case_1}
\end{figure}

Condition \eqref{eq:poly_local_condition_1} is violated in the
region in the immediate right of the interval $[a,b]$. This
implies that the absolute value of the evaluation of polynomial
function, $|p(x)|$, has to be greater than $k$ in the beginning of
the interval $[b, c]$. Also, we know that within $[a, b]$ and $[c,
d]$, $|p(x)|$ is strictly less than $k$. This can happen in two
ways, as shown in Fig.~\ref{fig:polynomial_case_1} (the other
possibility is that the polynomial can be the negative of that
shown in the figure, in which case the same argument holds). In
either of these ways, the function has to go through $-k$ value
twice in $[b,c]$. Therefore, by Rolle's theorem, $p^{\prime}(x_0)
= 0$, for some $b \leq x_0 \leq c$.

\begin{figure}[h]
  \centering
  \subfigure[Case 2(a)]{\label{fig:polynomial_3}\includegraphics[width=75mm]{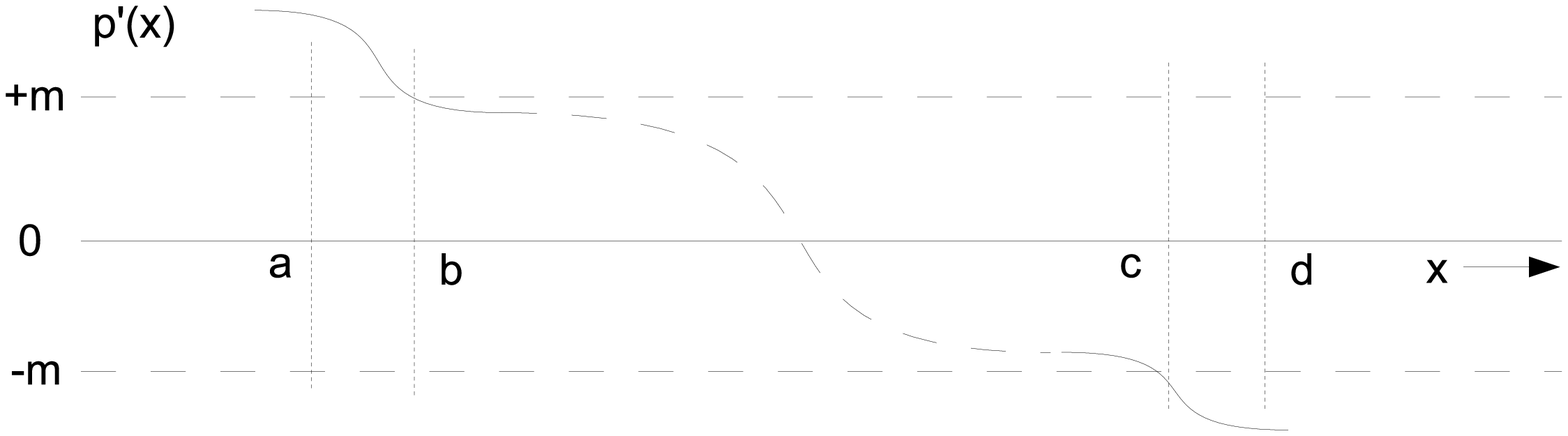}}
  \subfigure[Case 2(b)]{\label{fig:polynomial_4}\includegraphics[width=75mm]{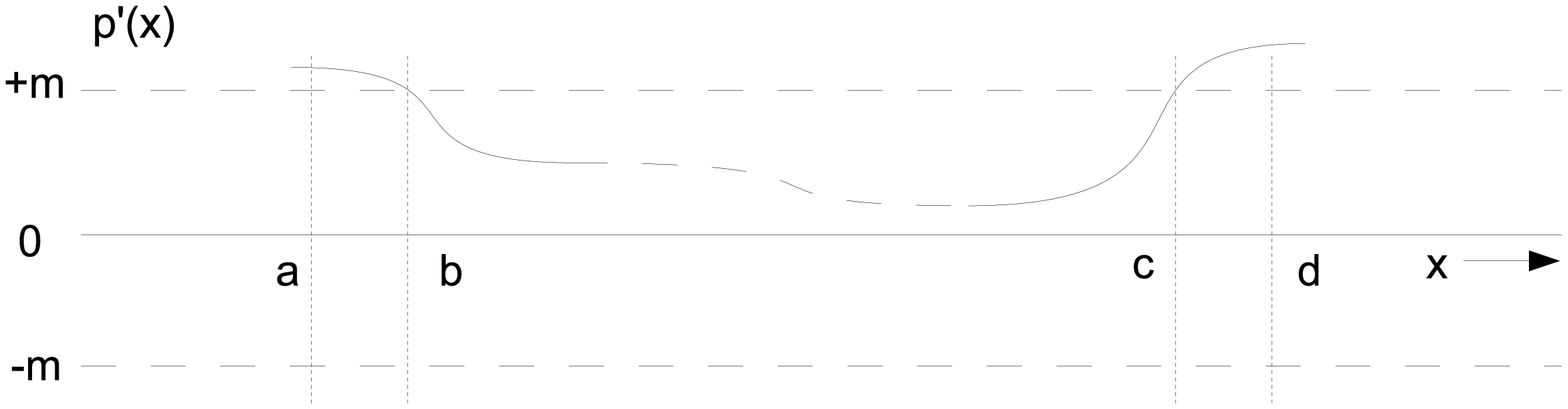}}
  \caption{Condition \eqref{eq:poly_local_condition_2} is violated in $[b, c]$. }
  \label{fig:polynomial_case_2}
\end{figure}

\emph{Case 2:} Condition \eqref{eq:poly_local_condition_2} is
violated at the end of interval $[a, b]$

This implies that the absolute value of the evaluation of the
derivative of $p(x)$, i.e.,$|p^{\prime}(x)|$, diminishes below $m$
in the beginning of the interval $[b, c]$.  Also, we know that
within $[a, b]$ and $[c, d]$, $|p^{\prime}(x)|$ is greater than or
equal to $m$. This can happen only in two ways, as shown in
Fig.~\ref{fig:polynomial_case_2}. In the first case,
$p^{\prime}(x_0) = 0$, for some $b \leq x_0 \leq c$. In the second
case, the function $p^{\prime}(x)$ takes the same value $+m$ twice
in $[b,c]$, and hence by Rolle's theorem, $p^{\prime\prime}(x_0) =
0$, for some $b \leq x_0 \leq c$.

By the above claim, for any two arbitrary intervals in ${\cal R}$,
there exists real root of $p^{\prime}(x)$ or $p^{\prime\prime}(x)$
between those two intervals. Since the number of roots of a
polynomial is bounded by its degree, there will be only finitely
many such intervals. In particular, the number of intervals $L$ is
bounded by $2d$, which is an upper bound on the total number of
zeros of $p(x)$ and $p^{'}(x)$.

\section{Proof of Lemma~\ref{lem:poly_probability} \label{app:poly_probability}}

 For any
polynomial $f$ in several gaussian random variables, we have that
\beqan \Pr \{ f(\bold{x}_1,\bold{x}_2,...,\bold{x}_N) \neq 0 \} &
= & 1. \eeqan This follows since letting
\beqan \underline{y} & = & (x_1,x_2,\cdots,x_{N-1}) \\
S & = & \{x_N \mid f(x_1,x_2,\cdots,x_N)=0\}, \eeqan we see that
\beqan \Pr \{ f(\bold{x}_1,\bold{x}_2,...,\bold{x}_N)
=  0 \} & = & \\
\int_{-\infty}^{\infty} \cdots \int_{-\infty}^{\infty}
f(\underline{y})  \int_{x_N \in S} f(x_N/\underline{y}) dx_N
d\underline{y} & = & 0 , \eeqan because the innermost integral
equals zero as $S$ is finite given a particular assignment of
\underline{y}, i.e., \beq \int_{x_N \in S} f(x_N/\underline{y})
dx_N \ = \ 0 . \eeq

Let $\bold{x} := \{ \bold{x}_1,\bold{x}_2,...,\bold{x}_N \}$. Let
us define an indicator function $I_{\delta}(\bold{x})$ as follows:
\beqan I_{\delta}(\bold{x}) & :=  &  \left\{ \begin{array}{ccc}
        1 ,& &  \ |f(\bold{x}_1,\bold{x}_2,...,\bold{x}_N)| < \delta \\
        0 ,& &  \ \text{else} \\
    \end{array}
    \right\}.   \eeqan
Then \beqa \lefteqn{\Pr \{ |f(\bold{x}_1,\bold{x}_2,...,\bold{x}_N)|^2
< \delta \}} \\ \label{eq:polynomial_bound}
 & = & \mathbb{E}_{\bold{x}} \ I_k(\bold{x}) \nonumber  \\
 & = & \mathbb{E}_{\bold{x}_1,\bold{x}_2,...,\bold{x}_{N-1}} \
 \mathbb{E}_{\bold{x}_N} \{ I_k(\bold{x}) \ | \ \bold{x}_1^{N-1} \} \nonumber \\
 & = & \mathbb{E}_{\bold{x}_1,\bold{x}_2,...,\bold{x}_{N-1}} \
\Pr  \ \{ |f({x}_1,{x}_2,...,{x}_{N-1},\bold{x}_N)|^2 < \delta\}.
\nonumber \\ \label{eq:poly_exp_vars}  \eeqa

Let $ f(\bold{x_N}) := f({x}_1,{x}_2,...,{x}_{N-1},\bold{x}_N)$,
where the dependence of $f$ on the first $N-1$ variables is made
implicit. Let \beqan f(\bold{x}_N) = \sum_{k=0}^{d_N} b_k
\bold{x}_N^k, \eeqan where $d_N$ is the degree of the polynomial
$f$ in the variable $x_N$.   Since $b_{d_N}$ is a polynomial in
the variables $x_1^{N-1}$, it follows from the lemma above that
with probability one, $b_{d_N} \neq 0$.

Let \beqan g(x_N) = \frac{\partial f(x_N)}{\partial x_N} \eeqan be
the partial derivative of $f(x_N)$ with respect to $x_N$.  Then we
can write
\beqa \lefteqn{\Pr  \ \{ |f({x}_1,{x}_2,...,{x}_{N-1},\bold{x}_N)| < \delta \}} \\
& = & \Pr \ \{ |f(\bold{x}_N)| < \delta, |g(\bold{x}_N)| \geq \delta^{1/2} \} \nonumber \\
& & + \Pr \ \{ |f(\bold{x}_N)| < \delta, |g(\bold{x}_N)| <
\delta^{1/2} \} \nonumber \\
& \leq & \Pr \ \{ |f(\bold{x}_N)| < \delta, |g(\bold{x}_N)| \geq \delta^{1/2} \} \nonumber \\
& & + \Pr \ \{ |g(\bold{x}_N)| < \delta^{1/2} \}.
\label{eq:poly_split_into_two} \eeqa

\begin{figure}[h]
  \centering
   \includegraphics[width=75mm]{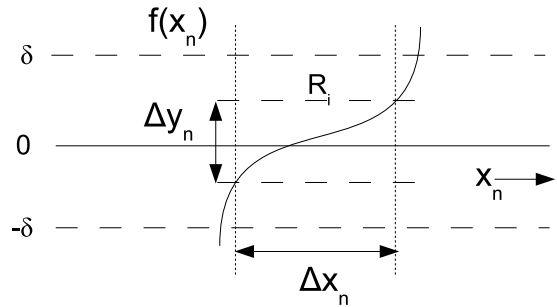}
   \caption{f(x) in a region $R_i$}
  \label{fig:derivative}
\end{figure}

Let us consider the first term on the RHS. The region ${\cal R} :=
\{ |f({x}_N)| < \delta, |g({x}_N)| \geq \delta^{1/2} \}$ is
described by two conditions $|f({x}_N)| < \delta$ and $|g({x}_N)|
\geq \delta^{1/2} $. It is shown in
Lemma~\ref{lem:poly_finite_regions} that the set of all values of
$x_N$ satisfying both conditions can be expressed as the union of
$L$ pairwise-disjoint intervals $R_i, i=1,2,...,L$ with $L <
2d_N$. Now $\Pr(x_N \in R) = \sum_{i=1}^{L} \Pr(\bold{x}_N \in
R_i)$. We will now proceed to upper-bound the probability
$\bold{x}_N \in R_i$. To do so, consider
Fig.~\ref{fig:derivative}. Let $\Delta x_n$ be the width of the
interval $R_i$ and $\Delta f(x_n)$ be the height (equal to the
difference in maximum and minimum values of $f(.)$ in $R_i$).
Since the slope of the curve $g(x)$ is greater than $\delta^{1/2}$
throughout $R_i$, we have that

\beqan \left | \frac{\Delta f(x_n)}{\Delta x_n} \right | & \geq &
\delta^{1/2}.\eeqan For our purposes, we can assume without loss
of generality that \beqan \frac{\Delta f(x_n)}{\Delta x_n}  & \geq
& \delta^{1/2}, \eeqan which gives us \beqan \Delta x_n & \leq &
\frac{\Delta f(x_n)}{\delta^{1/2}} \ . \eeqan

However in any contiguous region, $\Delta f(x_n) \leq 2 \delta$.
This implies that \beqan x_n & \leq & \frac{2
\delta}{\delta^{1/2}} = 2\delta^{1/2}. \eeqan Since $\bold{x}_n$
is a $\mathcal{N}(0,1)$ random variable, we have that $\Pr \{
\Delta \bold{x}_n \leq a \} \leq  c a$, where $c$ is the maximum
value of the gaussian pdf.

Therefore, \beqan  \{\bold{x} \in R_i \} & \subset &  \{
\bold{x}_n \leq 2\delta^{1/2}
\} \\
\Pr \{ \bold{x} \in R_i \} & \leq &  \Pr \{ \bold{x}_n \leq 2\delta^{1/2} \} \\
& \leq & 2c \delta^{1/2}. \eeqan  Using \beqan
 \Pr \{ \bold{x} \in R \} & = & \sum_{i=1}^{L} \Pr \{\bold{x} \in R_i\},  \eeqan we obtain  \beqa \Pr \ \{ |f({\bold{x}}_N)| <
\delta, |g({\bold{x}}_N)| \geq \delta^{1/2} \} & \leq & L 2c  \delta^{1/2} = C \delta^{1/2}. \nonumber \\
\label{eq:poly_first_term} \eeqa

Plugging \eqref{eq:poly_first_term} into
\eqref{eq:poly_split_into_two} yields
\beqa \Pr \ \{ |f({\bold{x}}_N)| < \delta \}  & \leq & C \delta^{1/2} + \Pr \ \{ |g({\bold{x}}_N)| < \delta^{1/2} \} . \nonumber \\
\label{eq:poly_order_recursion} \eeqa

Since $g(x)$ is of lower degree than $f(x)$, the process can be
continued to yield \beqa  \Pr \ \{ |f({\bold{x}}_N)| <
\delta \}  & \leq & C \delta^{1/2} + C \delta^{1/4} + ... + C \delta^{1/{2^{d_N-1}}} \nonumber \\
& &  + \Pr \{ d_N! \ {b}_{d_N} \leq \delta^{1/{2^{d_N}}} \}.
\label{eq:poly_one_var} \eeqa

Only the last term involving $b_{d_N}$ is a function of the
remaining variables $x_1,x_2,...,x_{N-1}$.  We next substitute
\eqref{eq:poly_one_var} into \eqref{eq:poly_exp_vars} and take the
expectation over the remaining variables.  This yields
\beqa \lefteqn{\Pr \{ |f(\bold{x}_1,\bold{x}_2,...,\bold{x}_N)| < \delta \}} \\
& \leq &  C \delta^{1/2} + C \delta^{1/4} + ... + C \delta^{1/{2^{d_N-1}}} \nonumber \\
& &  + \mathbb{E}_{\bold{x}_1,\bold{x}_2,...,\bold{x}_{N-1}} \
I_{\{ d_N! \ \bold{b}_{d_N} \leq \delta^{1/{2^{d_N}}} \}}.
\label{eq:poly_semifinal}\eeqa

The last term is identical to that in the right hand side of
\eqref{eq:polynomial_bound}, except that the polynomial $d_N! \
b_{d_N}$ involves $(N-1)$ or fewer variables and hence this
procedure can be continued.  Eventually, we will be left with the
probability that a constant coefficient $J$ is greater than
$\delta^{1/s}$ for some integer $s$. Choosing the constant $K$
appearing in the statement of the lemma to equal $J^s$, we obtain
that this probability is equal to the probability that $K \leq
\delta$. But by hypotheses, $K>\delta$ and hence this probability
is equal to zero. This allows us to rewrite the bound on
probability appearing in \eqref{eq:poly_semifinal} as
\beqan \Pr \{ |f(\bold{x}_1,\bold{x}_2,...,\bold{x}_N)| < \delta \}  \\
\ \leq  \ C_1 ( \delta^{1/2} + \delta^{1/4} + ... +
\delta^{1/{2^{e}}} ) && \eeqan for a suitable constant $C_1$ and
some integer $e$.

Choosing $K \leq 1$ forces $\delta < 1$ since by hypotheses,
$\delta < K$.    In this case \beqan \delta^{1/2} + \delta^{1/4} +
... + \delta^{1/{2^{e}}} & \leq  & e \delta^{1/{2^e}} \eeqan

With $A := eC_1$ and $d:=2^e$, we get \beqa \Pr \{
|f(\bold{x}_1,\bold{x}_2,...,\bold{x}_N)| < \delta \} & \leq & A
\delta^{1/d} \eeqa as desired.

\section{Proof of Lemma~\ref{lem:parallel_channel}\label{app:parallel_channel}}

Let ${\bf x}, {\bf y}, {\bf H}$ denote the concatenated input and
output vectors and channel matrix respectively, i.e., ${\bf x =
[x_1,x_2,\ldots,x_M]}^{T},{\bf y = [y_1,y_2,\ldots,y_M]}^{T}$ and

\beqa {\bf H} =  \left[\begin{array}{cccc}
    {\bf H}_1 & & & \\
     & {\bf H}_2 & &\\
     &  & \ddots &\\
     &  &  & {\bf H}_M \\
    \end{array}\right] . \eeqa

Then the input-output relation of the parallel channel is given by
${\bf y = Hx + w}$. We now proceed to determine the probability of
outage.  We have:
\beqa \lefteqn{I(\bold{x};\bold{y} | \bold{H} = H )} \nonumber \\
& = & h(\bold{y} | \bold{H} = H )- \sum_{i=1}^{M} h(\bold{y}_i | \bold{y}_1^{i-1}, \bold{x}, \bold{H} = H ) \nonumber \\
& = & h(\bold{y} | \bold{H} = H )- \sum_{i=1}^{M} h(\bold{y}_i | \bold{x}_i ,
\bold{H} = H ) \nonumber \\
& \leq & \sum_{i=1}^{M} h(\bold{y}_i | \bold{H} = H )-
\sum_{i=1}^{M} h(\bold{y}_i |
\bold{x}_{i} , \bold{H} = H ) \nonumber \\
& = & \sum_{i=1}^{M} [ h(\bold{y}_i | \bold{H} = H )- h(\bold{y}_i
| \bold{x}_{i} ,
\bold{H} = H ) ] \nonumber \\
& = & \sum_{i=1}^{M} [ h(\bold{y}_i | \bold{H}_i = H_i )-
h(\bold{y}_i | \bold{x}_i ,
\bold{H}_i = H_i ) ] \nonumber \\
& = & \sum_{i=1}^{M} I(\bold{x}_i;\bold{y}_i | \bold{H}_i = H_i ).
\eeqa

\beqan \Rightarrow \lefteqn{\Pr\{I(\bold{x};\bold{y} \mid \bold{H} = H) \leq r\log\rho\}} \\
& \geq & \Pr\{ \sum_{i=1}^{M} I(\bold{x}_i;\bold{y}_i | \bold{H}_i = H_i ) \leq r\log\rho \} \eeqan

Equality in the equation above holds if all the $\bold{x}_i$ are
independent. So we will choose the $\bold{x}_i$ to be independent,
for the rest of the discussion, since this maximizes the mutual
information and hence minimizes the outage probability. Define
$\bold{Z}_i := I(\bold{x}_i;\bold{y}_i | \bold{H}_i = H_i )$. Thus
$\bold{Z}_i$ is a function of the channel realization $H_i$ and is
therefore a random variable.  Since $\{ \bold{H}_i \}$ are
independent by the hypothesis of the lemma and $\bold{x}_i$ are
independent by the argument above, $ \{ \bold{Z}_i \} $ are also
independent.

Let $R = r \ \log(\rho)$ and $R_i = r_i \ \log(\rho)$ for
$i=1,2,\ldots,M$. Our next goal is to evaluate $\Pr \{ \
\sum_{i=1}^{M} \bold{Z}_i \leq r \log(\rho) \ \}$.    To do this,
we first consider the case when $M=2$ and we evaluate $\Pr \{ \
\bold{Z}_1 + \bold{Z}_2 \leq r \log(\rho) \ \}$. Then we extend
this to general $M$ by induction. We define
\beqan F_{Z_i} (R_i) & := & \Pr \{\bold{Z}_i < R_i \}  \ \\
f_{Z_i} (R_i) & := & \frac{d}{d{R_i}} F_{Z_i} (R_i) \\
\text{Let } F_{Z_i} (R_i) &  \doteq & \rho^{-d_i(r_i)} \\
\text{Then } f_{Z_i} (R_i) &  \doteq & \frac{d}{d{r_i \ \log(\rho)}} \rho^{-d_i(r_i)}\\
&  \doteq  & \rho^{-d_i(r_i)} (-1) \frac{d}{d{r_i}}
d_i(r_i) \\
&  \doteq  & \rho^{-d_i(r_i)} . \eeqan The last equation follows
since $d_i(r_i)$ is a decreasing function making $-\frac{d}{dr_i}$
positive.

\beqan \Pr(\bold{Z}_1+\bold{Z}_2 \leq R) & = & \rho^{-d(r)}\\
 & = & \int_{0}^{\infty} \ {f_{Z_1} (R_1) F_{Z_2} (R-R_1) d{R_1} } \\
 & \doteq & \int_{0}^{\infty} \ \rho^{-d_1(r_1)} \ \rho^{-d_2(r-r_1)} \log (\rho) d(r_1). \eeqan

By Varadhan's Lemma\cite{DemZei}, the SNR exponent of the integral
is given by

\beqan d(r) & = & \inf_{r_1 \geq 0} {d_1(r_1)} + d_2(r-r_1) \\
& = & \inf_{(r_1,r_2): \ r_1+r_2=r} \ \sum_{i=1}^{2} {d_i(r_i)}. \eeqan Proceeding by induction, we get for the
general case with $M$ parallel channels where \beqan \rho^{-d(r)} & \doteq & \Pr \{ \ \sum_{i=1}^{M} Z_i \leq r
\log(\rho) \ \} \eeqan that \beqan d(r) & = & \inf_{(r_1,r_2,\cdots,r_M): \ \sum_{i=1}^{M} r_i = r} \
\sum_{i=1}^{M} {d_i(r_i)}. \eeqan

\section{Proof of Lemma~\ref{lem:parallel_dependent} \label{app:parallel_correlated}}

Following the same line of reasoning as in the proof of Lemma~\ref{lem:parallel_channel}, we choose $\bold{x_i}$
to be independent. For computing the DMT, we know from Lemma~\ref{lem:signal_white} that the inputs can in fact
be independent and identically distributed with a $\mathbb{C}\mathcal{N}(0,I)$ distribution. So we have \beqan
\lefteqn{I(\bold{x};\bold{y} | \bold{H} = H )} \\
& & = \sum_{i=1}^{M}I(\bold{x}_i;\bold{y}_i | \bold{H}_i = H_i ) \eeqan \beqan \lefteqn{\Pr \{
I(\bold{x};\bold{y}
| \bold{H} = H ) \leq r \log{\rho} \}} \\
& & = \Pr \{\sum_{i=1}^{M} I(\bold{x}_i;\bold{y}_i | \bold{H}_i =
H_i ) \leq r \log{\rho} \} \eeqan \beqan & & = \Pr \{ \sum_{i=1}^N
\ n_iI(\bold{x}_i;\bold{y}_i | \bold{H}_i = H_i ) \leq r
\log{\rho} \} \eeqan

Now, define $Z_i := n_iI(\bold{x}_i;\bold{y}_i | \bold{H}_i = H_i )$. Also let
\beqan \rho^{-\tilde{d}_i(r)} & \doteq & \Pr \{Z_i < {r}\log(\rho) \} \\
& = & \Pr \{I(\bold{x}_i;\bold{y}_i | \bold{H}_i = H_i ) < \left(\frac{r}{n_i}\right)\log(\rho) \} \\
& = & \rho^{-d_i\left(\frac{r}{n_i}\right)} \\
\text{where, }\rho^{-d_i(r)} & \doteq & \Pr \{I(\bold{x}_i;\bold{y}_i | \bold{H}_i = H_i ) < r \log(\rho) \}.
\eeqan

Using the same convolution argument in the proof of Lemma~\ref{lem:parallel_channel}, \beqan d(r) & = &
\inf_{(r_1,r_2,\cdots,r_N): \ \sum_{i=1}^{N} r_i = r} \
\sum_{i=1}^{N} {  \tilde{d}_i(r)} \\
& = & \inf_{(r_1,r_2,\cdots,r_N): \ \sum_{i=1}^{N} r_i = r} \ \sum_{i=1}^{N} {d_i\left(\frac{r_i}{n_i}\right)} \\
& = & \inf_{(r_1,r_2,\cdots,r_N): \ \sum_{i=1}^{N} \ n_i r_i = r} \ \sum_{i=1}^{N} {d_i(r_i)} . \eeqan

\section*{Acknowledgment}

Thanks are due to K.~Vinodh and M.~Anand for useful discussions.

\begin{spacing}{1.2}

\end{spacing}

\end{document}